\PassOptionsToPackage{pdfpagelabels=false}{hyperref}
\documentclass[fleqn,usenatbib]{mnras}

\usepackage[T1]{fontenc}
\usepackage{ae,aecompl}

\usepackage{graphicx}	
\usepackage{amsmath}	
\usepackage{amssymb}	
\usepackage{mathptmx}
\usepackage{txfonts}
\usepackage{multicol}
\usepackage{dblfloatfix}
\usepackage[caption = false, position=t,singlelinecheck=off]{subfig}
\usepackage{reledmac}
\arrangementX[A]{twocol}
\usepackage{epstopdf}
\usepackage{cleveref}
\usepackage{color, soul}

\newcommand{\ddfrac}[2]{\frac{\partial #1}{\partial #2}}
\newcommand{\mmsn}[1]{$#1_\mathrm{MMSN}$}
\newcommand{\e}[1]{$e_\mathrm{#1}$}

\newcommand{\Pb}[1]{$#1\,P_\mathrm{b}$}
\crefname{figure}{Fig.}{Figs.}

\title[Self-Gravity in Circumbinary Systems]{The Role of Disc Self-Gravity in Circumbinary Planet Systems: \\ I. Disc Structure and Evolution}

\author[M. M. Mutter et al.]{
Matthew M. Mutter$^{1}$\thanks{E-mail: m.m.mutter@qmul.ac.uk\newline},
Arnaud Pierens$^{2,3}$ and
Richard P. Nelson$^{1}$
\\
$^{1}$Astronomy Unit, Queen Mary University of London, Mile End Rd, London, E1 4NS, UK\\
$^{2}$Universit\'e de Bordeaux, Observatoire Aquitain des Sciences de \v lUnivers, BP89 33271 Floirac Cedex, France\\
$^{3}$Laboratoire \v dAstrophysique de Bordeaux, Univ. Bordeaux, CNRS, B18N, all\'ee Geoffroy Saint-Hilaire, 33615 Pessac, France
}

\date{Accepted XXX. Received YYY; in original form ZZZ}

\pubyear{2016}

\begin{document}
\let\textlabel\label
\label{firstpage}
\pagerange{\pageref{firstpage}--\pageref{lastpage}}
\maketitle

\begin{abstract}
We present the results of 2-dimensional hydrodynamic simulations of self-gravitating circumbinary discs around binaries whose parameters match those of the circumbinary planet-hosting systems Kepler-16, -34 and -35. Previous work has shown that non-self-gravitating discs in these systems form an eccentric precessing inner cavity due to tidal truncation by the binary, and planets which form at large radii migrate until stalling at this cavity. While this scenario appears to provide a natural explanation for the observed orbital locations of the circumbinary planets, previous simulations have failed to match the observed planet orbital parameters. The aim of this work is to examine the role of self-gravity in modifying circumbinary disc structure as a function of disc mass, prior to considering the evolution of embedded circumbinary planets. In agreement with previous work, we find that for disc masses between 1--5 times the minimum mass solar nebula (MMSN), disc self-gravity affects modest changes in the structure and evolution of circumbinary discs. Increasing the disc mass to 10 or 20 MMSN leads to two dramatic changes in disc structure. First, the scale of the inner cavity shrinks substantially, bringing its outer edge closer to the binary. Second, in addition to the eccentric inner cavity, additional precessing eccentric ring-like features develop in the outer regions of the discs. If planet formation starts early in the disc lifetime, these changes will have a significant impact on the formation and evolution of planets and precursor material.
\end{abstract}




\section{Introduction}\label{sec:intro}

The discovery of the first circumbinary planet around a main-sequence short-period binary -- Kepler-16b by \citet{Doyle2011} -- further strengthened the growing sense that extrasolar planets are essentially ubiquitous. We now know of planets on a broad range of orbits around single stars of different spectral class and evolutionary stage \citep{Wolszczan1992, Mayor1995, Marois2008}, around one star within a binary star system \citep{Hatzes2003}, and around both members of close binary star systems. Since the announcement of Kepler-16b, the Kepler spacecraft has discovered a further 10 circumbinary planets: 
Kepler-34b and Kepler-35b\footnoteA{\citep{Welsh2012b}}, Kepler-38b\footnoteA{\citep{Orosz2012}}, Kepler-47b,c\footnoteA{\citep{Orosz2012a}}, \& d\footnoteA{\citep{Welsh2015a}}, Kepler-64b\footnoteA{\citep{Kostov2013}}, Kepler-413b\footnoteA{\citep{Kostov2014}}, Kepler-453b\footnoteA{\citep{Welsh2015}} and KIC-5473556\footnoteA{\citep{Kostov2015}}.

Even before their discovery it was recognised that this class of objects would be interesting for dynamical and planet formation studies. \citet{Holman1999} found that there exists a critical limit for dynamical stability around a short-period binary. This stability limit depends on the eccentricity of the binary and its mass ratio, as well as its semi-major axis. With the exception of the outer planets in Kepler-47, and KIC-5473556, all the Kepler circumbinary planets are found close to this critical limit. The tendency for these planets to lie very close to this region, as well as their inferred co-planarity with the binary's orbital plane, suggests that these planets formed in a common circumbinary disc and migrated in, before halting at their current stable locations.

In principle, these planets could also have formed in-situ from the material surrounding the binary, however this process must progress under the strong gravitational influence of the binary at these locations. The barrier lies in bringing sufficient numbers of planetesimals together, in a manner which results in mass growth, until a planet is formed. Previous studies have shown numerous disruptive effects to this goal: differential pericentre-alignment of eccentric planetesimals of different sizes leads to corrosive collisions \citep{Scholl2007}; gravitational interaction, with asymmetric features in the gas disc and global eccentric mode, leading to large impact velocities of planetesimals \citep{Marzari2008, Kley2010}; N-body simulations show excitation of planetesimal eccentricity leading to relative velocities disruptive to accretion and the formation of planetary bodies, possibly out to $\approx10$ au depending on system parameters \citep{Paardekooper2012, Meschiari2012, Meschiari2012a, Lines2014, Bromley2015}.

These issues, alongside the proximity of these planets to their host binary's stability limit and their coplanarity, suggest that they formed further out in the disc and moved into the their current positions by disc-driven migration -- either Type-I \citep{Ward1997, Tanaka2002} or Type-II \citep{Lin1986, Nelson2000}. Thus, the question of how these planets stopped at their current location needs answering. It turns out that this question had already been answered prior to the discovery of circumbinary planets. The influence of the binary on the circumbinary disc exerts a tidal torque which sweeps material away from the binary, creating a central cavity that can act as a barrier to migration. The radial extent of this cavity depends on the binary eccentricity, mass ratio and semi-major axis, as well as disc parameters \citep{Artymowicz1994}. An inner cavity in a circumbinary disc has been directly imaged around the GG Tau system \citep{Dutrey1994}. Subsequent studies show that the interaction of the binary with this feature lead to an asymmetric, eccentric, precessing disc \citep{Pierens2013, Pelupessy2013, Kley2014}.

The interaction of giant migrating planets with the circumbinary disc was first studied by \citet{Nelson2003}. This study showed that Jovian-mass planets generally migrated into the central cavity, where they were then captured into a 4:1 mean-motion-resonance with the binary. These giant planets often underwent close-encounters with the binary, with scattering events ejecting them from the system. Lighter, Saturnian-mass planets underwent stable migration to the disc cavity where they then stopped \citep{Pierens2008}. Less massive planets undergo Type-I migration until they are halted at the inner cavity edge by a strong positive corotation torque, counteracting the Lindblad torque -- see \citet{Pierens2007, Pierens2008, Pierens2008a} for more details. \citet{Dunhill2013}, using 3D SPH simulations, argued that the near-circular orbit of Kepler-16b is evidence that it formed in a massive disc where the disc material could significantly damp the orbit of the planet.

The techniques developed in these works were then applied to a number of the newly discovered Kepler circumbinary systems, in attempts to explain and recreate the orbits of their planets \citep{Pierens2013, Kley2014, Kley2015}. \citet{Pierens2013}, henceforth referred to as PN13, had difficulty recreating both the semi-major axis and eccentricity for the observed planets, with a range of disc parameters, under an isothermal equation of state. \citet{Kley2014, Kley2015}, referred to from now on as KH14 and KH15, included a more realistic equation of state and radiation effects, as well as the role of multi-planet migration and interaction. These works also had difficulty in recreating all the observed properties of the Kepler circumbinary planets. What is now clear is that understanding how the disc evolves, and what physics shape its structure, are key to explaining the current state of this exotic class of planets.

To this end we aim to investigate the impact of self-gravity, and by extension disc mass, on the evolution and structure of circumbinary discs as well as the evolution of planets in these discs. Self-gravity has already been examined in low-mass circumbinary discs \citep{Marzari2009}, where it was discounted as an unimportant factor in the disc evolution. However, as pointed out in \citet{Lines2015}, even at low-mass, disc self-gravity can modify the precession frequencies associated with low-frequency global eccentricity modes \citep{Papaloizou2002}. The modest influence of self-gravity in circumbinary discs is in contrast to its apparent importance in determining the evolution of circumprimary discs \citep{Marzari2009}. While self-gravity has little impact at low disc mass, its influence becomes apparent at higher mass, modifying the physical size of the tidally truncated cavity and leading to the formation of large scale features in the circumbinary disc \citep{Lines2015}. In this paper, and a follow-up paper (Mutter et al, in prep., hereafter referred to as Paper II) we aim to investigate the effect of self-gravity over a large range of disc masses and binary systems. The motivation is to probe the early dynamical history of circumbinary discs -- as we increase the disc mass we examine earlier and earlier times in the system's history. We aim to address the questions: Does a high-mass disc leave a fingerprint on the planet population if circumbinary planets form early, or is this erased by the transition to a low-mass disc as the system evolves? Does the epoch when planets form, accrete gas, and migrate affect their final orbital configuration or mass? 

In this paper we look at a number of different phenomena in systems that are intended to mimic the Kepler-16, -34 and -35 circumbinary systems. The first issue that we address is the treatment of the inner boundary of the computational domain, including the outflow condition used, the radius of the inner boundary and embedding the binary within the computational domain. This investigation is motivated by similar discussions in \citet{Marzari2009}, PN13, KH14 and \citet{Lines2015}, which have found a range of different disc outcomes for a range of inner boundary conditions. Using a more physically realistic prescription for the treatment of the inner boundary we obtain a set of evolved discs with masses corresponding to 1, 2, 5, 10 and 20 times the MMSN model \citep{Hayashi1981} in their respective systems. We examine their evolution and the structure induced by their self-gravity. These discs will then we used to examine the evolution of planets at different times in the discs' history in Paper II.

The outline for this paper is as follows. Section 2 outlines the physical model and initial conditions used in the simulations. Section 3 summarises our investigation into boundary condition issues, as well as attempts to model the inner boundary and disc more accurately. Section 4 examines the evolution and final structure of self-gravitating circumbinary discs in the Kepler-16, -34 and -35 systems. Our results from Sections 3 and 4 are summarised, and discussed in Section 5.

\section{Numerical Setup}\label{sec:num_setup}
Although the numerical and hydrodynamical model used in this study are similar to previous work (see \citet{ Pierens2007, Pierens2008, Pierens2008a}, PN13), there are a number of significant departures or modifications. For clarity, the following section is a detailed run-through of the parameters and initial conditions used.
\subsection{Equations of Motion}\label{sec:eq_of_mot}
\subsubsection{Disc Evolution}\label{sec:disc_evol}
The equations of motion of the binary-disc-planet system are detailed here in two-dimensional polar co-ordinates $(R,\,\phi)$ with the origin kept at the center of mass of the binary. 
Under the two-dimensional approximation we consider vertically integrated quantities, such as the disk surface density $\Sigma = \int^{\infty}_{-\infty}\rho \mathrm{d}z$.
In this set-up the continuity equation is:
\begin{equation}
	\ddfrac{\Sigma}{t} + \nabla\,\mathbf{\cdot}\,(\Sigma \mathbf{v}) = 0
\label{eq:continuity}
\end{equation}
The equations for the evolution of the radial and the angular momentum are given by:
\begin{equation}
	\ddfrac{\Sigma \upsilon_{R}}{t} + \nabla\,\mathbf{\cdot}\,(\Sigma \upsilon_{R}\mathbf{v}) - \frac{\Sigma \upsilon_\phi^2}{R} = -\ddfrac{p}{R} - \Sigma \ddfrac{\Phi}{R} + f_R
\label{eq:v_rad}
\end{equation}
and
\begin{equation}
	\ddfrac{\Sigma \upsilon_\phi}{t} + \nabla\,\mathbf{\cdot}\,(\Sigma \upsilon_\phi\mathbf{v}) + \frac{\Sigma \upsilon_R \upsilon_\phi^2}{R} = -\frac{1}{R}\ddfrac{p}{\phi} - \frac{\Sigma}{R} \ddfrac{\Phi}{\phi} + f_\phi
\label{eq:v_phi}
\end{equation}
Here, $p$ is the vertically-averaged pressure, $(\upsilon_{R},\,\upsilon_\phi)$ are the radial and azimuthal components of the gas velocity $\mathbf{v}$ , $f_R$ and $f_\phi$ are the components of the vertically-averaged viscous force per unit volume (described in \citet{Nelson2000}). 
For the specific simulations presented in this paper, consisting of a close binary system on a fixed orbit surrounded by a self-gravitating disc, the gravitational potential experienced by the disk can be split into the following terms:
\begin{equation}
\Phi = \Phi_\mathrm{SG} +  \sum_{k=1}^{2}\Phi_{\mathrm{s},k},
\label{eq:pot}
\end{equation}
where $\Phi_{\mathrm{s},k}$ is the direct potential from the $k^\text{th}$ star of the binary, with mass $M_{\mathrm{s},k}$:
\begin{equation}
	\Phi_{\mathrm{s},k} = 
	\begin{cases}
		-\frac{G M_{\mathrm{s},k} \left(3R_{\mathrm{Roche},k}^2 - d^2\right)}{2 R_{\mathrm{Roche},k}^3}, &  \text{for}\ d < R_{\mathrm{Roche}, k} \\
		-\frac{G M_{\mathrm{s},k}}{d}, & \text{if}\ \ \ d \geqslant R_{\mathrm{Roche}, k}.
	\end{cases}
\label{eq:pot_star}
\end{equation}
Here $ d = |\mathbf{R^\prime} - \mathbf{R}_{\mathrm{s},k}|$ and $R_{\mathrm{Roche},k}$ is the radius of the $k^\text{th}$ star's Roche lobe, as defined in \citet{Eggleton1983},
\begin{equation}
R_{\mathrm{Roche},1} = A\frac{0.4q_1^\frac{2}{3}}{0.6q_1^\frac{2}{3} + ln(1 + q_1^\frac{1}{3})},
\label{eq:roche_lobe}
\end{equation}
where $A$ is the orbital separation between the two binary stars, and $q_1 = M_1/M_2$.
This prescription effectively treats the members of the binary as constant-density solid-body spheres with sizes equal to their respective Roche lobes. 
The first term in Eq. \ref{eq:pot}, $\Phi_\mathrm{SG}$, arises from the disc self-gravity and takes the form:
\begin{equation}
\Phi_\mathrm{SG} = \sum_i^{N_R} \sum_j^{N_\phi}  \frac{G m_{\mathrm{c},ij}} {|\mathbf{R_{ij}|}},
\label{eq:pot_SG}
\end{equation}
where $\mathbf{R_{ij}} = \mathbf{R}_{i} - \mathbf{R}_{j} + \epsilon_\mathrm{SG}$. $\epsilon_\mathrm{SG}$ is a softening length used to avoid singularities in the calculation of the disc's potential; it takes a value equal to 0.4$H$ in this work, where $H$ is the disc thickness.
The quantity $m_{\mathrm{c},ij}$ is the mass contained within cell $ij$, calculated with the surface density $\Sigma_{c,ij}$ and the surface area of the cell $A_{ij}$: 
\begin{equation*}
m_{\mathrm{c},ij} = \Sigma_{\mathrm{c},ij}\, A_{ij}.
\end{equation*}
The disc potential causes two additional acceleration terms -- $g_R$ and $g_\phi$ -- and are given in Appendix A of \citet{Baruteau2008}.

As mentioned already, the disc potential described above applies to a binary system that remains on a fixed orbit throughout its evolution and where the centre of mass of the binary system remains fixed in inertial space. For a binary that evolves in time due to the forces exerted on it by the disc, it is customary to work in a frame centred on the binary centre of mass and to include additional indirect terms in the disc potential to account for the acceleration of the binary centre of mass. We do not consider this situation here because our focus is on understanding the response of the disc to the potential generated by binary systems with well defined orbital elements that match those of the Kepler circumbinary planets. For experimental purposes, however, we can also consider a situation where the binary system maintains fixed orbital elements, but where the disc accelerates the centre of mass of the binary such that the centre of mass between the disc and the binary system is preserved. This can be achieved by including the indirect term in the disc potential due to the disc accerelating the binary centre of mass (while working in a frame that is centred on the binary centre of mass), but without evolving the orbital elements of the binary system. We undertake such a test calculation in section~\ref{sec:sg_origin} to demonstrate that our results are not influenced strongly by working with a binary system whose centre of mass is fixed in inertial space.

\subsubsection{Orbital Evolution}\label{sec:orb_evol}
Table \ref{tab:kepler_params} contains the best-fit observed binary and planetary orbital and mass parameters of the Kepler-16, -34 and -35 circumbinary planetary systems, as quoted in \citet{Doyle2011} and \citet{Welsh2012b}, which will be used as the central binaries for all the work presented here. These specific systems are chosen, not only because the planetary system's parameters are so well defined, but they span a wide range of binary orbital configurations: Kepler-16 is a low eccentricity, non-unity mass ratio binary; Kepler-34 is a high eccentricity, unity mass ratio binary; and Kepler-35 is a low eccentricity, unity mass ratio binary.
As the eventual goal of this work is to recreate the observed state of the Kepler circumbinary systems, the binaries' orbital parameters remain fixed at their current observed values. In other words the binary members only interact with each other, not with the disc (or planet when included). It is worth noting that, as found in previous work (see \citep{Pierens2007, Kley2015, Fleming2016}), if permitted there will be a back-reaction from the disc onto the binary. This can lead to orbital evolution of the binary, including shrinkage of $a_\mathrm{b}$ and growth of \e{b}. We are aware that for the most massive discs presented here, these changes could be significant -- an issue we hope to revisit in a later paper. The binary configuration when the planets form and migrate in the disc, may also be significantly different to their current state. Interaction with the disc, and planet, will alter the initial parameters of the system. Acknowledging these issues, the binary orbital elements remain at their initialised, observed values however. The resulting equation of motion for the stars is as follows:
\begin{equation}
	\frac{d^2\mathbf{R}_{\mathrm{s},k}}{dt^2} = -\frac{GM_{\mathrm{s},l}(\mathbf{R}_{\mathrm{s},k} - \mathbf{R}_{\mathrm{s},l})}{|\mathbf{R}_{\mathrm{s},k} - \mathbf{R}_{\mathrm{s},l}|^3} 
\label{eq:motion_star}
\end{equation}
for the two stars $k$ and $l$.
\begin{table}
	\centering
	\caption{Binary and planet parameters.}
	\label{tab:kepler_params}
	\begin{tabular}{cccc}
		\hline
		\hline
		 & Kepler-16 & Kepler-34 & Kepler-35\\
		\hline
		$M_\mathrm{A}\ (M_\odot)$ & 0.690 & 1.048 & 0.888\\
		$M_\mathrm{B}\ (M_\odot)$ & 0.203 & 1.021 & 0.809\\
		$m_\mathrm{p}\ (M_J)$ & 0.333 & 0.220 & 0.127\\
		$q_\mathrm{b} = M_\mathrm{B}/M_\mathrm{A}$ & 0.294 & 0.974 & 0.912\\
		$q = m_\mathrm{p}/M_\ast$ & $3.54\times 10^{-4}$ & $1.01\times 10^{-4}$ & $7.13\times 10^{-5}$ \\
		$a_\mathrm{b}$ (au) & 0.224 & 0.228 & 0.176\\
		$a_\mathrm{p}$ (au) & 0.705 & 1.090 & 0.603\\
		\e{b} &	 0.159 & 0.521 & 0.142 \\
		\e{p} &	 0.007 & 0.182 & 0.042\\
		Reference &	\citep{Doyle2011} & \multicolumn{2}{c}{\citep{Welsh2012b}}  \\
		\hline
	\end{tabular}
\end{table}

\subsection{Hydrodynamic Model}\label{sec:hydro_model}
The simulation work-load for this project was split across two separate numerical codes, after comparing test simulations to verify that the results agreed. Wherever possible, the same set-up is used across both codes.
The first code we use is an upgraded version of the code {\small FARGO} \citep{Masset1999}, called {\small FARGO-ADSG} which includes the calculation of disc self-gravity as well as an adiabatic equation of state \citep{Baruteau2008b,Baruteau2008}. The second code used is {\small GENESIS}, a code which uses an advection scheme based on the \citet{VanLeer1977} monotonic transport algorithm to solve the disc equations. The version of {\small GENESIS} used has the {\small FARGO} time stepping upgrade, as well as a module to calculate self-gravity. In both codes the binary and planetary orbits are evolved using a fifth-order Runge-Kutta integrator scheme \citep{Press1992}. 

These codes were used to run 2D hydrodynamic simulations in the plane of the binary's orbit. The calculations presented here use a grid resolution of $N_R \times  N_\phi  = 550 \times 550$ cells. The radial grid-spacing is logarithmic, as required by the self-gravity calculations. This has the added benefit of having a finer grid in the inner region of the disc closest to the binary.

The following computational units are used: the total mass of the binary $M_\star = M_\mathrm{A} + M_\mathrm{B} = 1$, the gravitational constant $G = 1$ and the radius $R = 1$ is equivalent to 1 au. To present the results of simulations we use the binary orbital period, $P_\mathrm{b} = 2\pi \sqrt{GM_\ast/a^3_\mathrm{b}}$, as the unit of time.

\begin{figure}
	\centering
	\includegraphics[width = 0.45\textwidth]{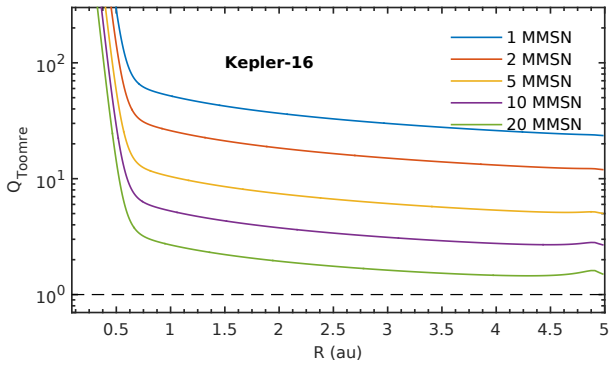}
	\caption{Radial profiles of the Toomre parameter corresponding to initial conditions in discs with masses corresponding to 1 -- \mmsn{20}. We can see that even the most massive discs satisfy the condition $Q > 1$ to remain stable, initially.}
	\label{fig:toomreq}
\end{figure}

\subsection{Initial Conditions}\label{sec:init_cond}
In initialising the disc surface density we follow the procedure in PN13 by setting the initial surface density to be:
\begin{equation}
\Sigma(R) = f_\mathrm{gap}\mathcal{X}\Sigma_0R^{-3/2},
\end{equation}
where $\Sigma_0$ is defined such that 2\% of the mass of the binary is contained within 30 au, $\mathcal{X}$ is a simple scaling factor used in Section \ref{sec:sg_res}, and $f_\mathrm{gap}$ is a gap-function used to simulate the inner cavity, decreasing the time needed for the disc to reach equilibrium. This function is taken to be, from \citet{Gunther2004}:
\begin{equation}
f_\mathrm{gap} = \left(1+\mathrm{exp}\left[-\frac{R - R_\mathrm{gap}}{0.1R_\mathrm{gap}}\right]\right)^{-1},
\label{eq:sigma}
\end{equation} 
where $R_\mathrm{gap} = 2.5a_\mathrm{b}$, the estimated gap size created by the tidal interaction of the binary on the disc \citep{Artymowicz1994}.

Building on the results of PN13 we choose disc parameters that match the disc models which most closely recreated the Kepler systems. To simulate the anomalous viscosity in the disc, an alpha model prescription \citep{Shakura1973} of effective kinematic viscosity is used, $\nu = \alpha c_s H$, where $\alpha = 10^{-3}$ is used throughout, $c_s$ is the sound speed in the disc and $H$ is the disc thickness.

We consider a disc with a constant aspect ratio, $h = \frac{H}{R} = 0.05$. This corresponds to a locally isothermal equation of state, with a temperature profile given by $T = T_0R^{-1}$. A more realistic equation of state, including radiative cooling, heating and radiation transfer etc., was not employed as we want to explore the impact of each major disc physics addition separately.

These initial conditions, with $\Sigma_0$ in Eq. \ref{eq:sigma} set to a value corresponding to a 1 $\times$ MMSN disc, lead to a minimum Toomre parameter value of $\approx$ 25 at the outer edge of the disc (see Fig.~\ref{fig:toomreq}). It is worth noting that the rotation profile in evolved circumbinary discs is non-Keplerian. Therefore whilst calculating the Toomre parameter, the assumption that the epicylic frequency, $\kappa$, is equal to the orbital frequency in the disc, $\Omega$, cannot be used. A more general formulation of the epicyclic frequency is used:
\begin{equation}
	\kappa^2 \equiv \frac{2\Omega}{R}\dfrac{d}{dR}\left(R^2\Omega\right).
	\label{eq:epicyclic_freq}
\end{equation}
The Toomre parameter at the inner edge of the disc is very high due to the low surface densities enforced by the gap function in Eq. \ref{eq:sigma}. This value of $Q > 1$ allows us to safely assume that this disc will be stable against gravitational fragmentation. Modifying $\mathcal{X}$ to higher values than 1, as per our method in later sections (see Section \ref{sec:sg_res}), this minimum value of $Q$ decreases, as the disc mass increases. In the most massive model, corresponding to a 20 MMSN disc, $Q \approx 1.5$ at the outer radius. 

\subsection{Boundary Conditions}\label{sec:bound_cond}
The computational domain lies between $R_\mathrm{in}$ and $R_\mathrm{out}$. A value of $R_\mathrm{out} = 5$ is used throughout this work. In Section \ref{sec:BC_res} the simulations presented use $R_\mathrm{in} = 1.5a_\mathrm{b}$, where $a_\mathrm{b}$ is the binary semi-major axis, whereas in later sections a different value is used -- discussed in Section \ref{sec:varrin_res}. 
A range of different boundary conditions are used, at either the inner or outer radii to govern how the material in the disc behaves at the radial edges:
\begin{itemize}
\item Closed -- or reflecting boundary condition. No flow of material is allowed across the disc edge. Material in the first (or last) active cell has its radial velocity set to 0.
\item Open -- material is allowed to freely leave the disc i.e. outflow. No inflow is allowed. A zero-gradient condition is set in both $\upsilon_R$ and $\Sigma$ (please note that this is for the case of outflow only).
\item Viscous -- this is a limiting condition to stop the inner disc from emptying of gas too quickly. Material in the innermost cells is given a radial velocity, $ \upsilon_R = \beta \upsilon_R(R_\mathrm{in}) $, where $\upsilon_R(R_\mathrm{in}) = -3\nu/2R_\mathrm{in}$ is the viscous drift velocity and $\beta$ is a free factor \citep{Pierens2008}. We follow previous works which use this condition and set $\beta = 5$.
\end{itemize}
We also need to define how the azimuthal velocity is set at the inner and outer boundaries. Usually in hydrodynamical codes the viscous stress is maintained by setting $\upsilon_\phi$ to the sub-Keplerian orbital velocity at the locations $R_\mathrm{in}$ and $R_\mathrm{out}$. However at the inner edge the potential created by the binary is extremely non-Keplerian. We therefore set a zero-gradient condition for the azimuthal velocity at this location -- $\upsilon_\phi$ in the first active cell takes the value of the second active cell.

A full investigation into the impact of the boundary condition at the inner edge is discussed in Section \ref{sec:rin_res}, with the last subsection highlighting our efforts to develop a more accurate method to deal with the inner edge of the disc.

\section{Treatment of the Inner Boundary}\label{sec:rin_res}

How best to model the inner boundary in hydrodynamic simulations of circumbinary discs has been examined a number of times (\citet{Marzari2009}, PN13, KH14 and \citet{Lines2015}). The condition imposed on the radial velocity -- Closed, Open or Viscous outflow -- has been shown to affect the structure of the inner disc, including the size of the tidally truncated cavity. It is unclear from these past results which boundary condition best models the transfer of mass and angular momentum in the inner disc, and therefore gives the most accurate rendering of the inner disc structure. For very eccentric binaries, and discs, material may be removed from the simulation at an Open boundary, when in fact it would have re-entered the disc \citep{Marzari2009}. Modelling this flow of material goes hand-in-hand with obtaining a physically accurate inner disc and cavity -- it has previously been shown that gas can flow across the cavity and onto the binary \citep{Artymowicz1996, Gunther2002}. As it has been shown to affect the final stopping position of migrating protoplanets, modelling the inner disc region correctly is key in any discussion of the evolution of circumbinary planets.

\subsection{Closed vs Open vs Viscous}\label{sec:BC_res}
To answer this question we have undertaken a systematic investigation into the impact of inner boundary condition choice on disc evolution in the three Kepler systems chosen. The simulations in this subsection adopt an inner radial boundary location equal to $1.5 a_\mathrm{b}$, such that the binary system orbits interior to the computational domain as has been done in previous work. The simulations were run until the disc structures achieved quasi-steady state.

\begin{figure}
	\centering
	\subfloat[]{\includegraphics[width = 0.39\textwidth]{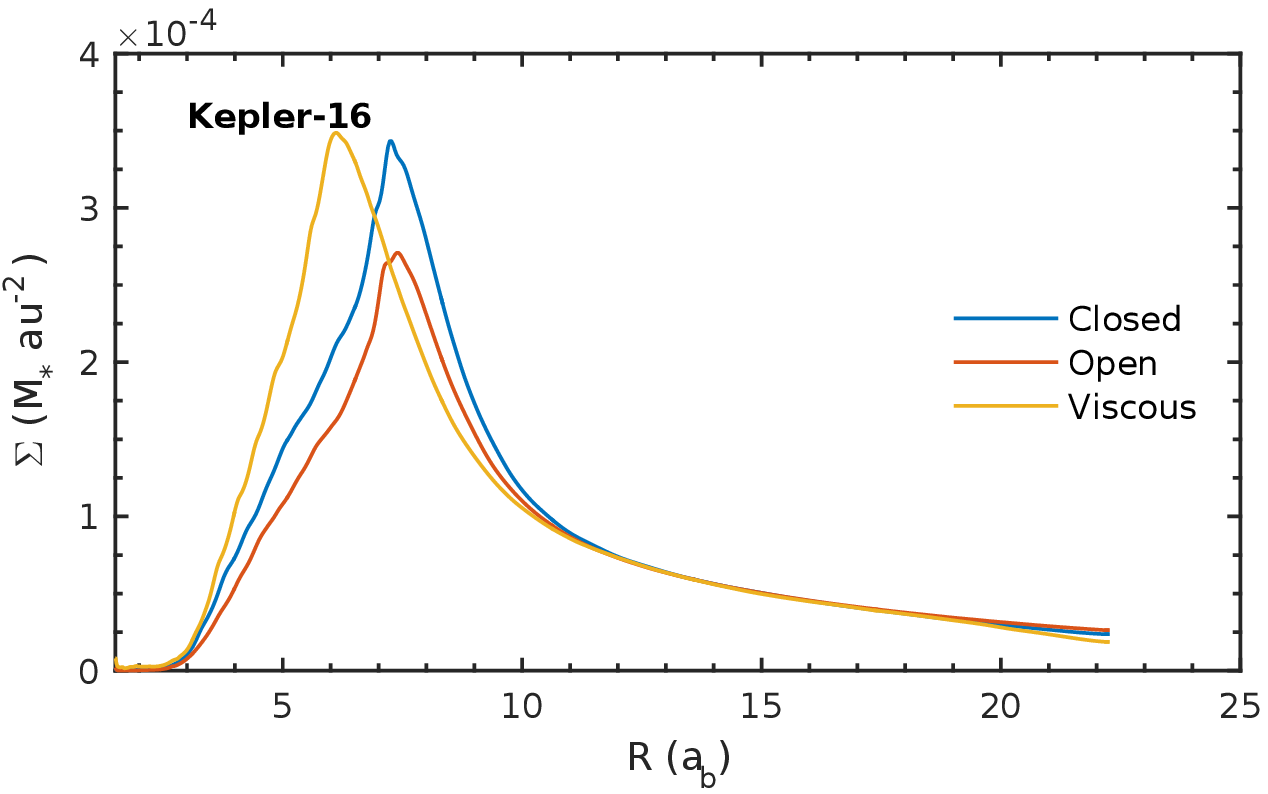}}\\
	\vspace{-10pt}
	\subfloat[]{\includegraphics[width = 0.39\textwidth]{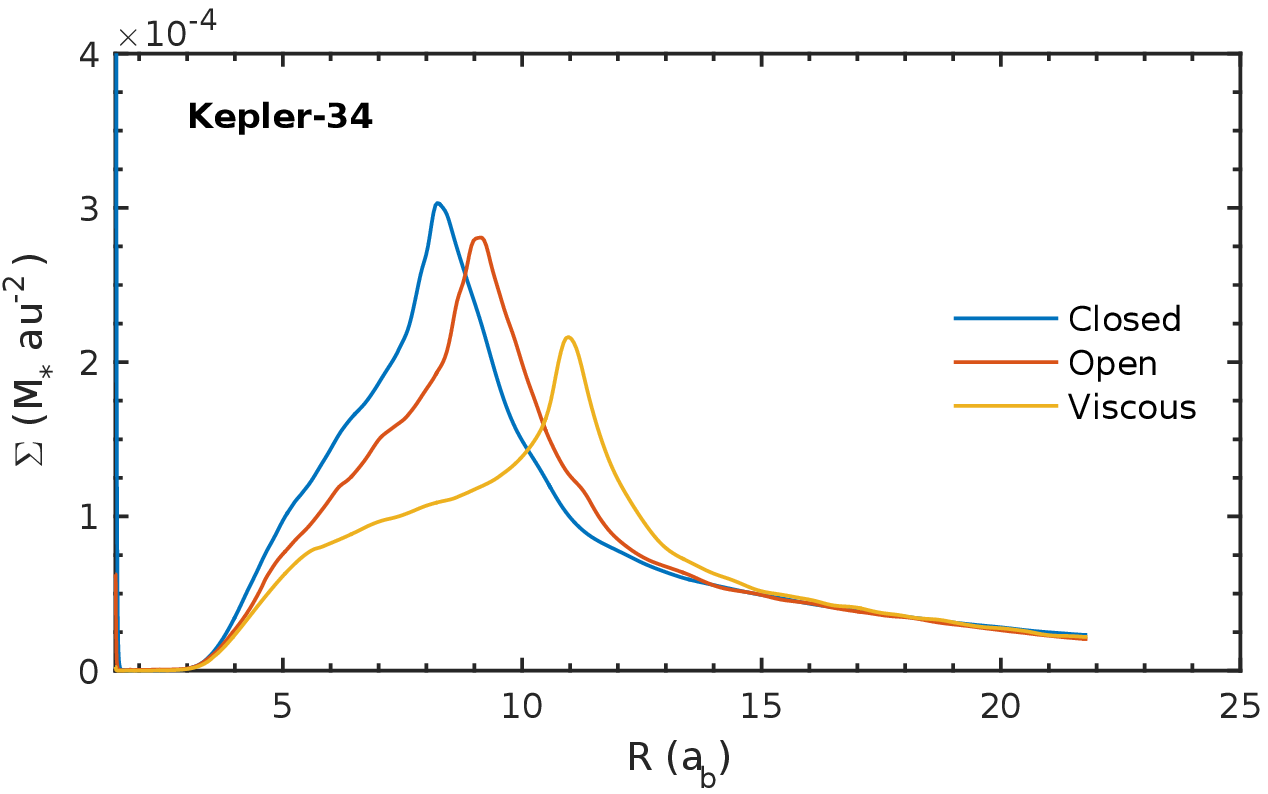}}\\
	\vspace{-10pt}
	\subfloat[]{\includegraphics[width = 0.39\textwidth]{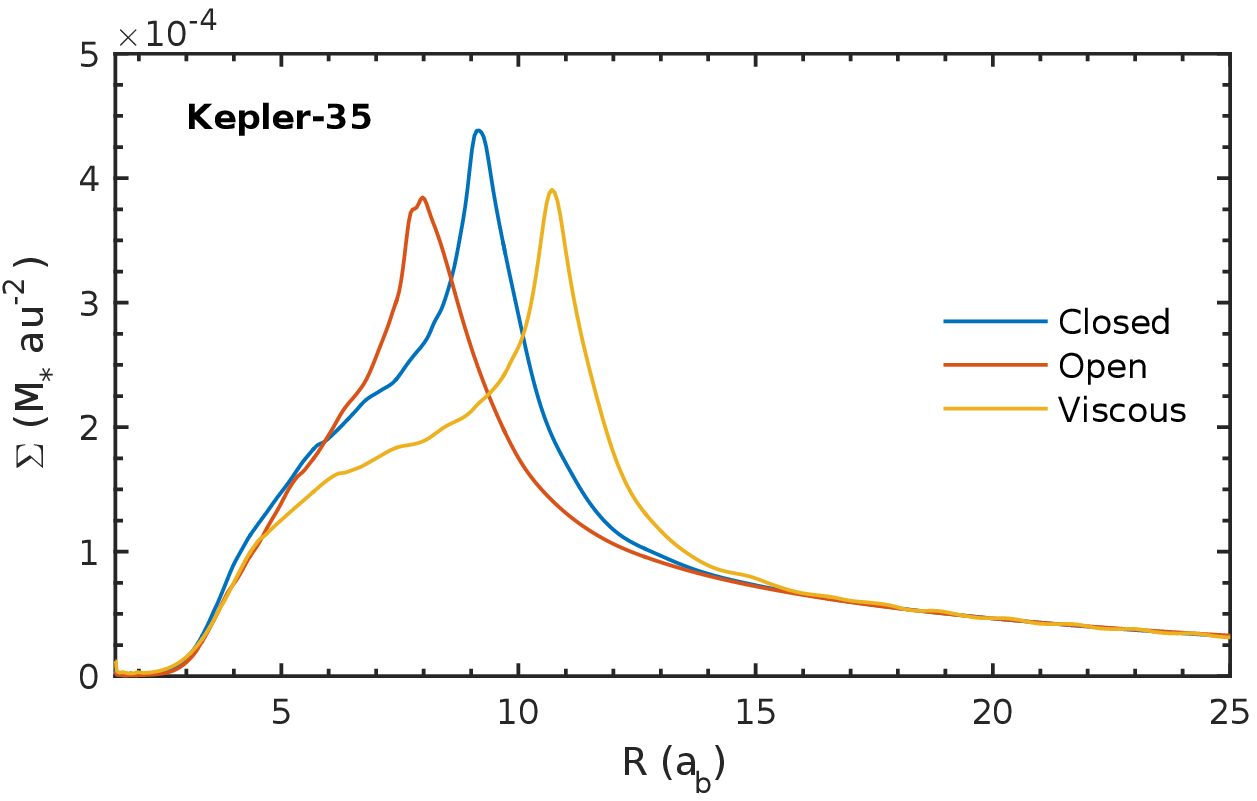}}\\
	\caption{Azimuthally averaged surface density profiles for the Closed, Open and Viscous conditions in the three systems. The disc radii are scaled by the respective binary semi-major axes in each system -- 0.224, 0.228 and 0.176 au for the Kepler-16, -34 and -35 systems. The inner disc radius lies at $1.5a_\mathrm{b}$ in these simulations.}
	\label{fig:bc_sd}
\end{figure}

The disparity in results obtained when varying the boundary choice is clear in Fig. \ref{fig:bc_sd}, with clear differences arising in peak surface density, and the radial extent of the eccentric inner cavity. We define the size of the cavity as the radial position of the peak surface density ($\Sigma_\mathrm{max}$) in the azimuthally-averaged surface density profile -- what we call $R_\mathrm{max}$. 

When comparing across the different systems, there are few common results for a given boundary condition and a rather confusing picture emerges. In the Kepler-16 system the maximum surface density is obtained with a Viscous boundary condition, whilst the largest eccentric cavity occurs in the Open case. In both the Kepler-34 and -35 systems a large eccentric cavity can be seen in the Viscous simulations, with the highest $\Sigma_\mathrm{max}$ achieved in Closed discs. In these two systems, each with a binary mass ratio close to unity, there is a large range in final disc configurations; both large and small inner cavities can been seen, with a difference of $\approx 50\%$ between the cavity sizes in the Closed and Viscous Kepler-34 systems.

The cavity sizes obtained from these simulations are in poor agreement with those obtained in PN13 and KH15. We attribute this to differences in how the gas velocities are treated at the inner and outer boundaries (PN13), and the lack of an enforced mass-flow through the disc (KH15) -- our discs are free to lose mass at their outer boundary.

\begin{figure*}
	\centering
	\subfloat[]{\includegraphics[width = 0.44\textwidth]{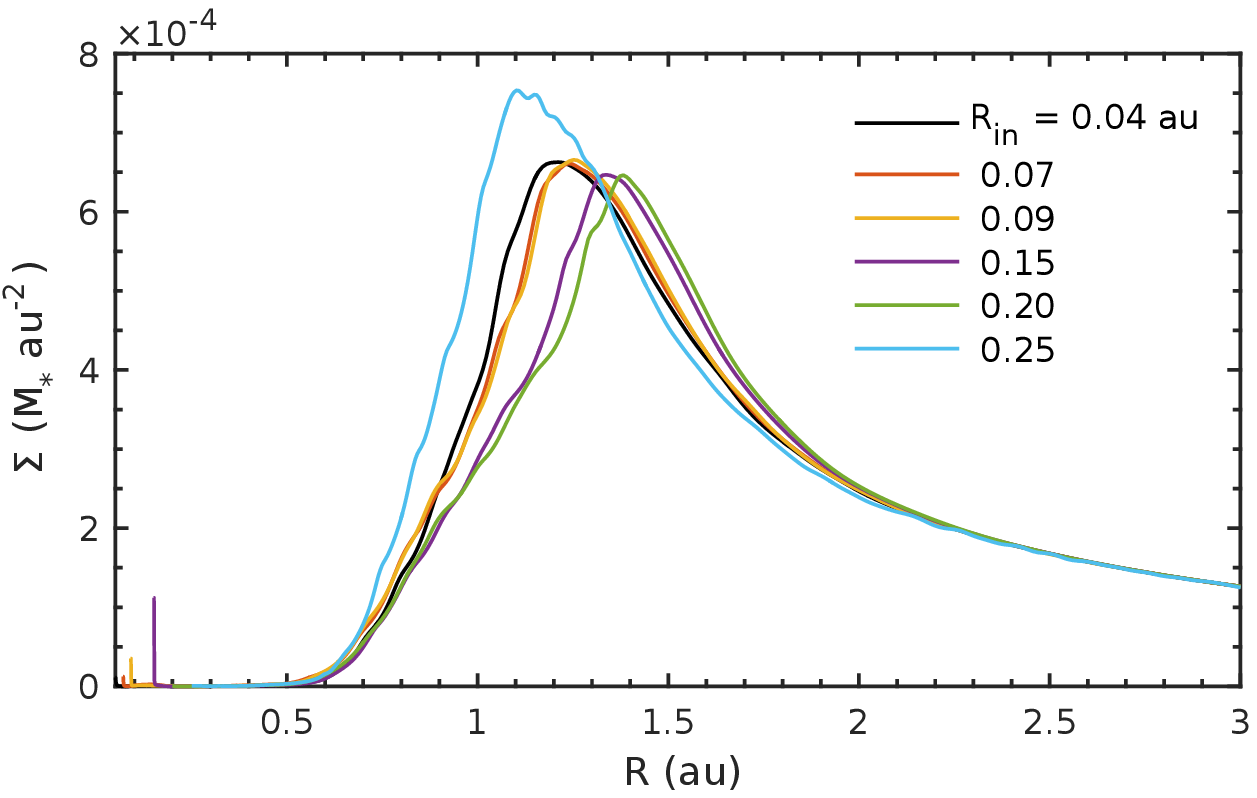}\label{fig:varrin_sd}}
	\hfil
	\subfloat[]{\includegraphics[width = 0.44\textwidth]{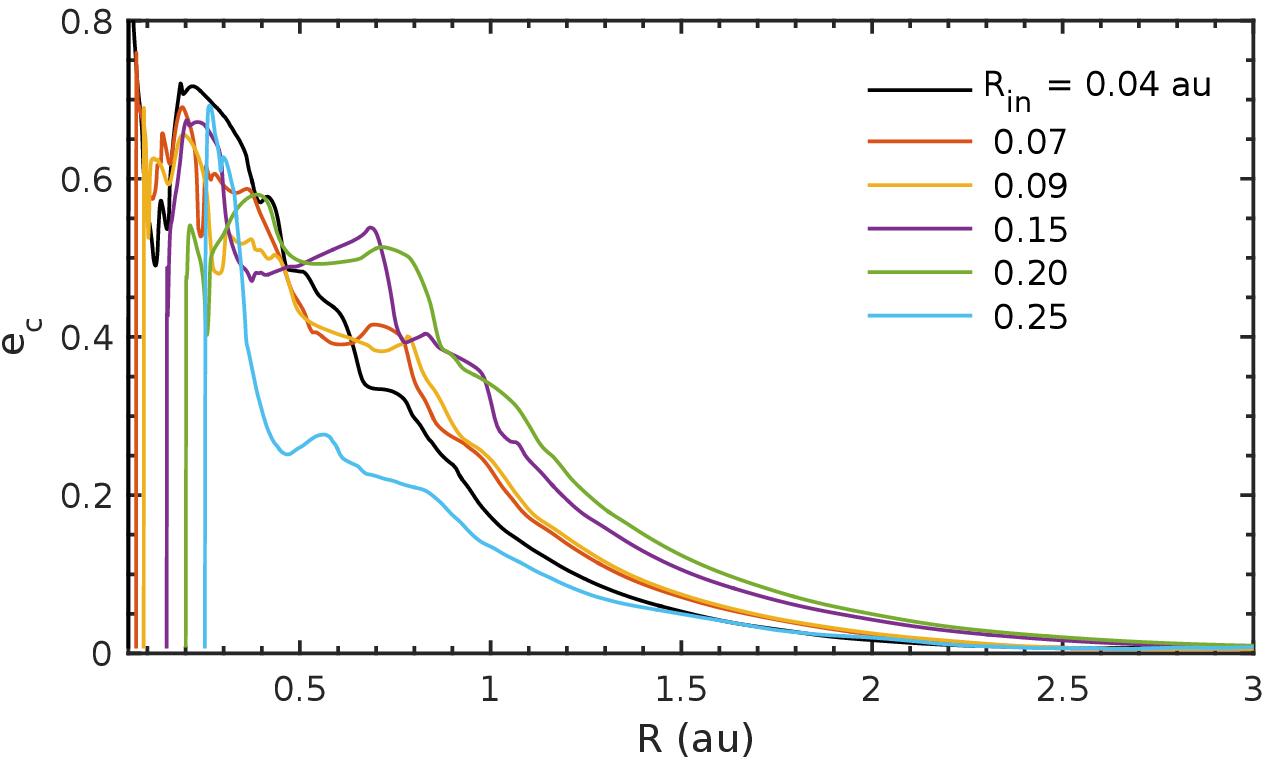}\label{fig:varrin_e}}
	\caption{\textit{a.)} Surface density distributions of increasing $R_\mathrm{in}$ boundary condition simulations in the Kepler-16 system at steady-state. The general trend for the eccentric inner cavity to increase in size as the inner radius increases can be seen. \textit{b.)} Azimuthally-averaged cell eccentricity profiles of increasing $R_\mathrm{in}$ boundary condition simulations in the Kepler-16 system at steady-state. The increasing size of the eccentric cavity can also be seen.}
	\label{fig:varrin_sd_e}
\end{figure*}

Similar differences are also seen in the azimuthally-averaged eccentricity profiles, with a range of results seen throughout the different discs, as well as the temporal evolution; some boundary conditions approach equilibrium at a faster rate, possibly due to over- or under-estimating the flow of material from the inner disc onto the binary.

The origins of the discrepancies in simulation outcomes as a function of the inner boundary condition are uncertain, and it would probably require an exhaustive and painstaking study in order to provide a convincing explanation. Such a study goes beyond the scope of this paper, the prime focus of which is exploration of the effects of self-gravity. Nonetheless, we expect that the steady-state disc structure (in some averaged sense) will arise because of a balance between the viscous evolution of the discs and the tidal torques due to the binary. The amplitudes of spiral density waves excited at Lindblad resonances, and their propagation into the disc and non-linear dissipation determine how the outward angular momentum flux from the binary is deposited into the orbiting disc material. The local viscous evolution also depends on local gradients in the disc, such that the steady profile achieved by the disc is expected to depend on the numerical set up to some extent. What is perhaps surprising is the observed strength of the dependency. 

It is worth noting that the above issues relating to boundary conditions pertain to codes that employ fixed Eulerian meshes. Lagrangian, particle-based codes such as SPH \citep{Artymowicz1994}, do not suffer from this issue as they implicitly adopt free boundary conditions, although issues relating to controlling numerical viscosity near the cavity edge and low particle numbers within the cavity do start to arise.

\subsection{Shrinking Inner Radius}\label{sec:varrin_res}

The lack of agreement between at least two boundary conditions in each system, motivated an investigation into a more realistic treatment of the inner disc. The choice of boundary condition tries to mimic the flow of material out of the disc and onto the central binary. Setting the inner boundary radius within the predicted extent of the cavity, but completely outside the orbit of the binary, requires a guess at how much material is really exchanged between the binary and the disc across the boundary. To completely capture this effect, the most realistic approach would be to completely embed the binary within the disc domain, an expensive procedure computationally as this would require small time steps for numerical stability.

By examining a suite of simulations of the Kepler-16 system, using ever smaller inner boundary radii -- from $0.25$ au down to a completely embedded case, $0.04$ au -- we aim to find a set-up which optimises accuracy and run-time, to use in the main body of this work. Whilst the number of radial grid cells is kept constant for all the different radial domain sizes shown here, we are confident that we are operating at a resolution high enough that issues caused by differences in the radial cell size are negligible. In terms of the local scale height at the inner boundary, our radial cell size ranges from between 0.25 scale heights and 0.16 scale heights, for the smallest and largest inner radii respectively. The same Open outer boundary condition at 5 au was used again, although the plots shown here are zoomed in to show the inner disc region more clearly.

\begin{figure}
	\includegraphics[width = 0.43\textwidth]{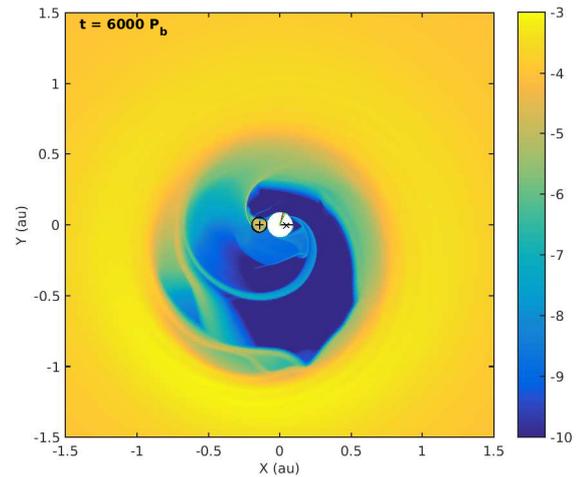}
	\vspace{-20pt}
	\caption{Log-scale plot of surface density of a Kepler-16 binary embedded in the $R_\mathrm{in} = 0.09$ au disc. Mass flow can be seen across the cavity through so-called streamer channels. This behaviour is not properly resolved by a larger inner disc radii.}
	\label{fig:sg_16_hight}
\end{figure}

\begin{figure}
	\centering
	\includegraphics[width = 0.43\textwidth]{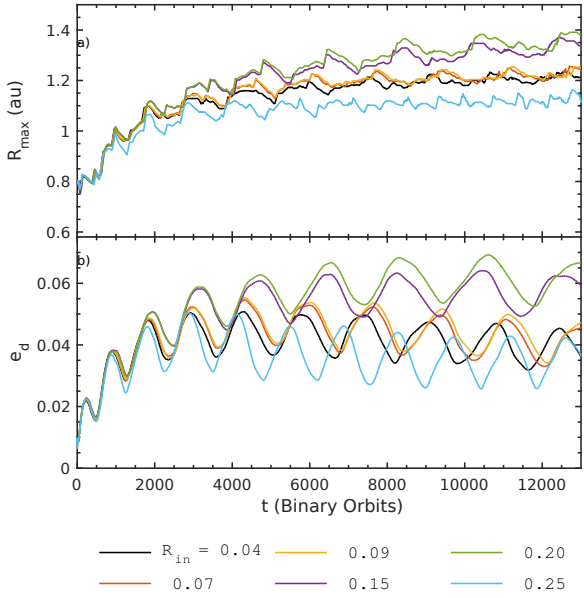}
	\caption{\textit{a.)} Evolution of $R_\mathrm{max}$ with increasing $R_\mathrm{in}$ in the Kepler-16 system, showing convergence to the embedded case as the inner radius shrinks. \textit{b.)} Disc-averaged eccentricity evolution of increasing $R_\mathrm{in}$ boundary condition simulations in the Kepler-16 system.}
	\label{fig:varrin_multi}
\end{figure}

As $R_\mathrm{in}$ increases, the size of the inner cavity (Fig. \ref{fig:varrin_sd}), and consequently $R_\mathrm{max}$ (top panel of Fig. \ref{fig:varrin_multi}) also increases. It follows from increasing the cavity size, that the extent of excited cell eccentricity and overall disc eccentricity should increase -- this can be seen in Figs. \ref{fig:varrin_e} and the bottom panel of \ref{fig:varrin_multi}. The cell eccentricity is calculated by treating each grid-cell as a particle, with the mass and velocity of the cell, orbiting the central binary. We then use the method in PN13 to calculate the disc-integrated eccentricity:
\begin{equation}
e_d = \frac{\int^{2\pi}_0 \int^{R_\mathrm{out}}_{R_\mathrm{in}}\Sigma e_\mathrm{c} R\, \mathrm{d}R\,\mathrm{d}\phi}{\int^{2\pi}_0 \int^{R_\mathrm{out}}_{R_\mathrm{in}}\Sigma R\,\mathrm{d}R\,\mathrm{d}\phi},
\label{eq:e_d}
\end{equation}
where $R_\mathrm{in}$ and $R_\mathrm{out}$ are the inner and outer radii of the disc, $\Sigma$ is the grid element surface density and \e{c} is the eccentricity of the cell-as-a-particle. Note that when calculating the disc eccentricity we treat each cell as if it were a ballistic particle orbiting around a single star containing the mass of the binary, and neglect the influence of the disc self-gravity on the calculation of the eccentricity. We note that the influence of self-gravity can be included \citep{Marzari2009}, but we have neglected this for simplicity as the magnitude change in the calculated disc eccentricity is small.
The instantaneous longitude of pericentre can also be calculated for each cell, $\omega_\mathrm{c}$, and a disc weighted average calculated in a similar manner:
\begin{equation}
\omega_d = \frac{\int^{2\pi}_0 \int^{R_\mathrm{out}}_{R_\mathrm{in}}\Sigma \omega_\mathrm{c} R\, \mathrm{d}R\,\mathrm{d}\phi}{\int^{2\pi}_0 \int^{R_\mathrm{out}}_{R_\mathrm{in}}\Sigma R\,\mathrm{d}R\,\mathrm{d}\phi}.
\label{eq:w_d}
\end{equation}

However, when calculating $\omega_\mathrm{d}$ we only include the regions of the disc which show non-negligible \e{c} values -- following KH15 -- neglecting material on circular orbits in the outer disc. A global disc calculation of $\omega_\mathrm{d}$ shows libration around a small range of angles, whereas it is clear from observing 2D surface density plots that the eccentric inner disc precession shows circulation. KH15 also showed that the local calculation agrees well with the position angle of the surface density peak, obtained from 2D density plots. This peak often lies near the apocentre of the eccentric material forming the cavity wall. This angle is obviously $180^\circ$ removed from the associated pericentre of the material.

Examining the results from these runs, we decided to choose $R_\mathrm{in} = 0.09$ au as the inner boundary radius for the Kepler-16 runs presented in the remainder of this article. It provides a speed-up over the completely embedded $0.04$ au case, but as can be seen in Figs. \ref{fig:varrin_sd_e} and \ref{fig:varrin_multi}, gives similar results. This corresponds to a disc where the Roche lobe of the lighter companion is always fully embedded in the active disc, and the primary star lies interior to the inner radius of the computational domain. Figure \ref{fig:sg_16_hight} shows the surface density distribution around and in the cavity surrounding the central binary once quasi-steady state has been reached. This snapshot was taken when the binary is at closest approach. The Roche lobe of the secondary star is shown in black -- clearly embedded in the disc. The white circle represents the area lying interior to the inner boundary of the computational domain. The cross and star symbols represent the positions of the primary and secondary stars, respectively.

Due to the masses of each the binary stars being equal in the Kepler-34 and -35 systems, the outer edge of the Roche lobes (i.e. the L1 Lagrange point) lie in close proximity to $R=0$. It is clear that in this situation, where the binary mass ratio is close to unity, a simulation with its origin at the centre of mass cannot also have one star embedded in the disc and the the other sitting interior to the boundary of the computational domain, and the procedure adopted for the Kepler-16 system cannot be used. Instead, an inner radius is chosen such that $> 70\%$ of the smallest Roche lobe is embedded in the disc at $r_\mathrm{min}$, corresponding to $R_\mathrm{in} = 0.04$ and $0.056$ au for the Kepler-34 and -35 systems respectively. These choices for $R_\mathrm{in}$, and the grid resolution chosen for these simulations, results in cells which are moderately elongated in the azimuthal direction compared to the radial direction. While this is not untypical for simulations of the type presented here, it does introduce truncation errors that are formally not equal in the $R$ and $\phi$ directions.
Even with the binary system partially embedded in the disc an outflow boundary condition is still needed at the inner edge; for all simulations presented from now on the limiting Viscous outflow condition was used. Whilst testing this disc set-up we found that for this reduced inner boundary radius, the difference to disc structure caused by either an Open or Viscous outflow condition was negligible.

\section{Self-Gravitating Discs}\label{sec:sg_res}

Using the disc set-up discussed in the previous sections and the boundary conditions summarised in Table \ref{tab:bc}, a set of runs examining the role of increasing the disc mass on the structure and dynamics of the circumbinary discs in the Kepler-16, -34 and -35 systems were undertaken. As the initial disc mass of the system increases we effectively simulate the circumbinary environment at earlier moments in its lifetime. The disc masses used correspond approximately to scaled-up versions of the Minimum Mass Solar Nebula (MMSN) model; $\Sigma_0$ is chosen so that $0.02\, \mathrm{M}_\odot$ is contained interior to 30 au. This $1_\mathrm{MMSN}$ model is then scaled by factors of 2, 5, 10, and 20 (refer to Fig. \ref{fig:toomreq} for radial profiles of the Toomre parameter in these discs). The three binary systems examined in this work all have different total masses, so we use a Minimum Mass \textit{Stellar} Nebula -- where $1_\mathrm{MMSN}$ contains $2\%$ of the mass of the central binary, $\mathrm{M}_\star$. We present each binary system in turn in our discussion of results below, whilst also splitting each set of runs into two disc families -- Low-mass (\mmsn{1}, \mmsn{2} and \mmsn{5}) and High-mass (\mmsn{10} and \mmsn{20}) discs.

\begin{table}
	\centering
	\caption{Inner and Outer Boundary Conditions}
	\label{tab:bc}
	\begin{tabular}{cccc}
		\hline
		\hline
		 & Kepler-16 & Kepler-34 & Kepler-35\\
		\hline
		$R_\mathrm{in}$ (au) & 0.090 & 0.040 & 0.056\\
		$R_\mathrm{in}$ BC & \multicolumn{3}{c}{Viscous}\\
		$R_\mathrm{out}$ (au) & \multicolumn{3}{c}{5.0}\\
		$R_\mathrm{out}$ BC & \multicolumn{3}{c}{Open}\\
		\hline
	\end{tabular}
\end{table}

\subsection{Kepler-16}\label{sg_16}

\begin{figure}
	\centering
	\includegraphics[width = 0.43\textwidth]{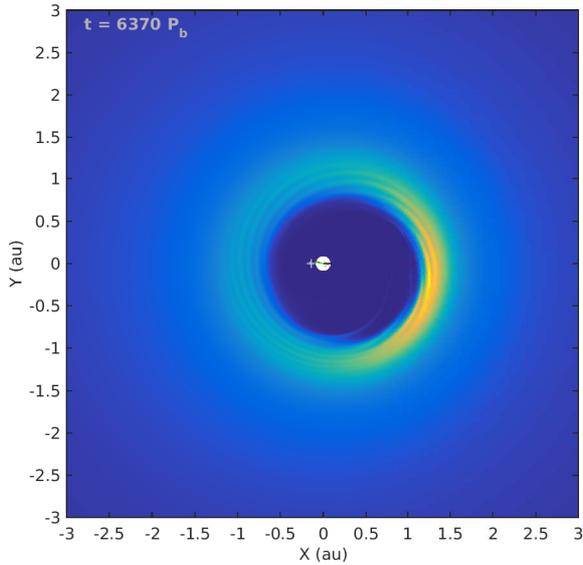}\label{fig:sg_16_sd1}
	\caption{Surface density plots of the Kepler-16 \mmsn{1} system at pseudo-steady-state. The characteristic eccentric inner cavity is clear with an edge at $\approx 1.2$ au, in good agreement with previous non-SG results.}
	\label{fig:sg_16_1}
\end{figure}

\begin{figure}
	\centering
	\subfloat[]{\includegraphics[width = 0.44\textwidth]{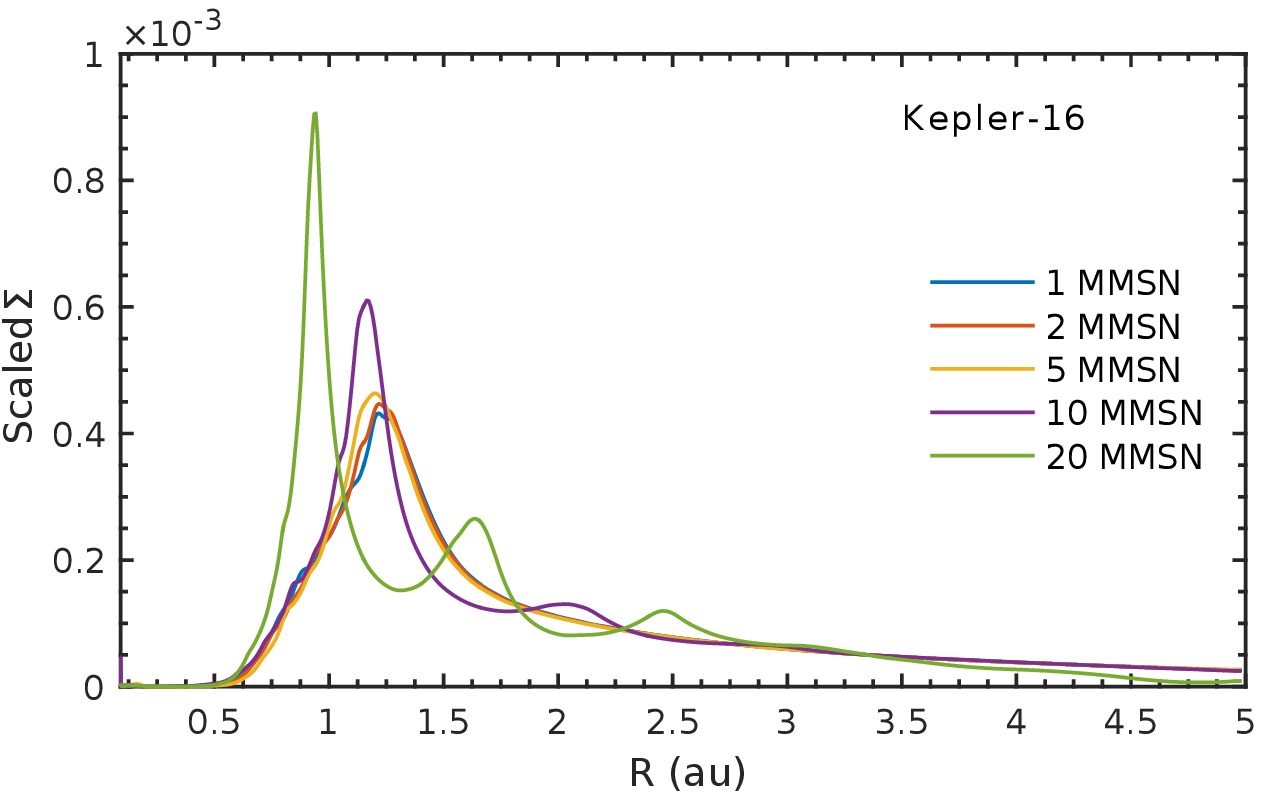}\label{fig:sg_16_sd}}\\
	\vspace{-13pt}
	\subfloat[]{\includegraphics[width = 0.44\textwidth]{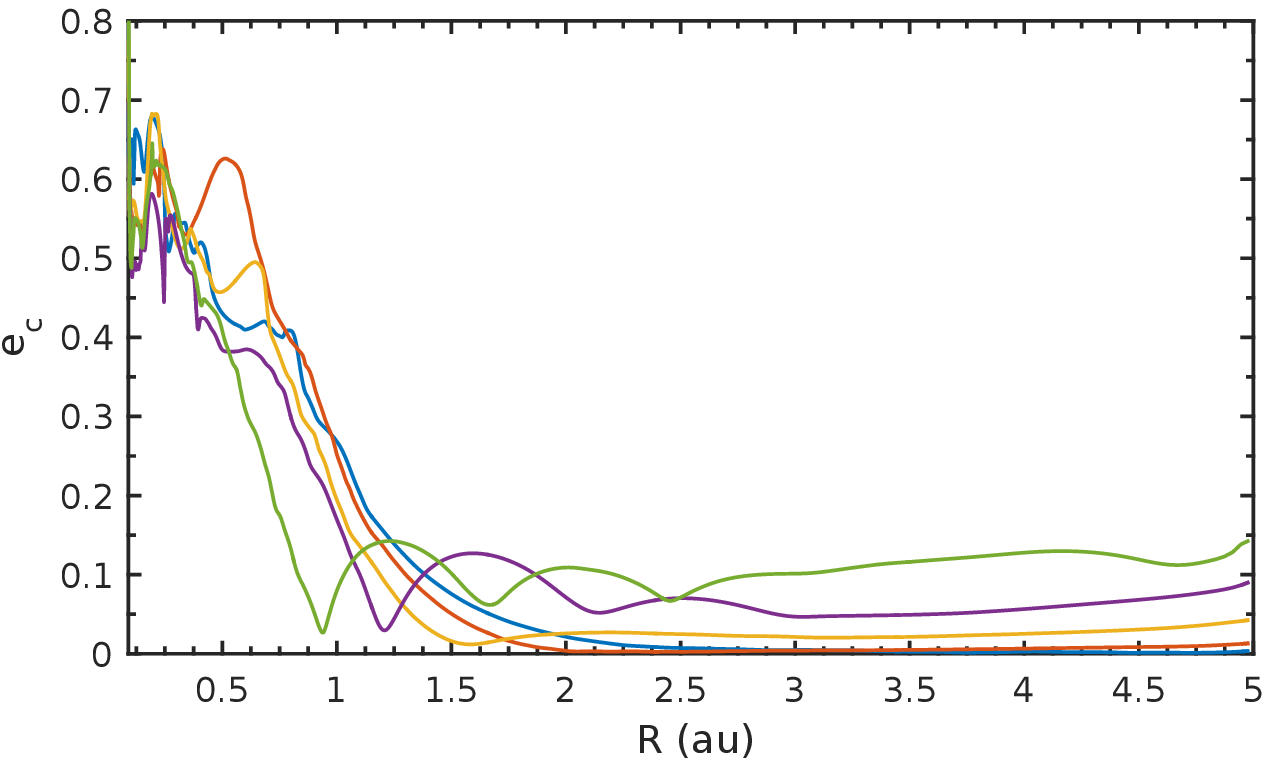}\label{fig:sg_16_e}}\\ 
	\vspace{-13pt}
	\subfloat[]{\includegraphics[width = 0.455\textwidth]{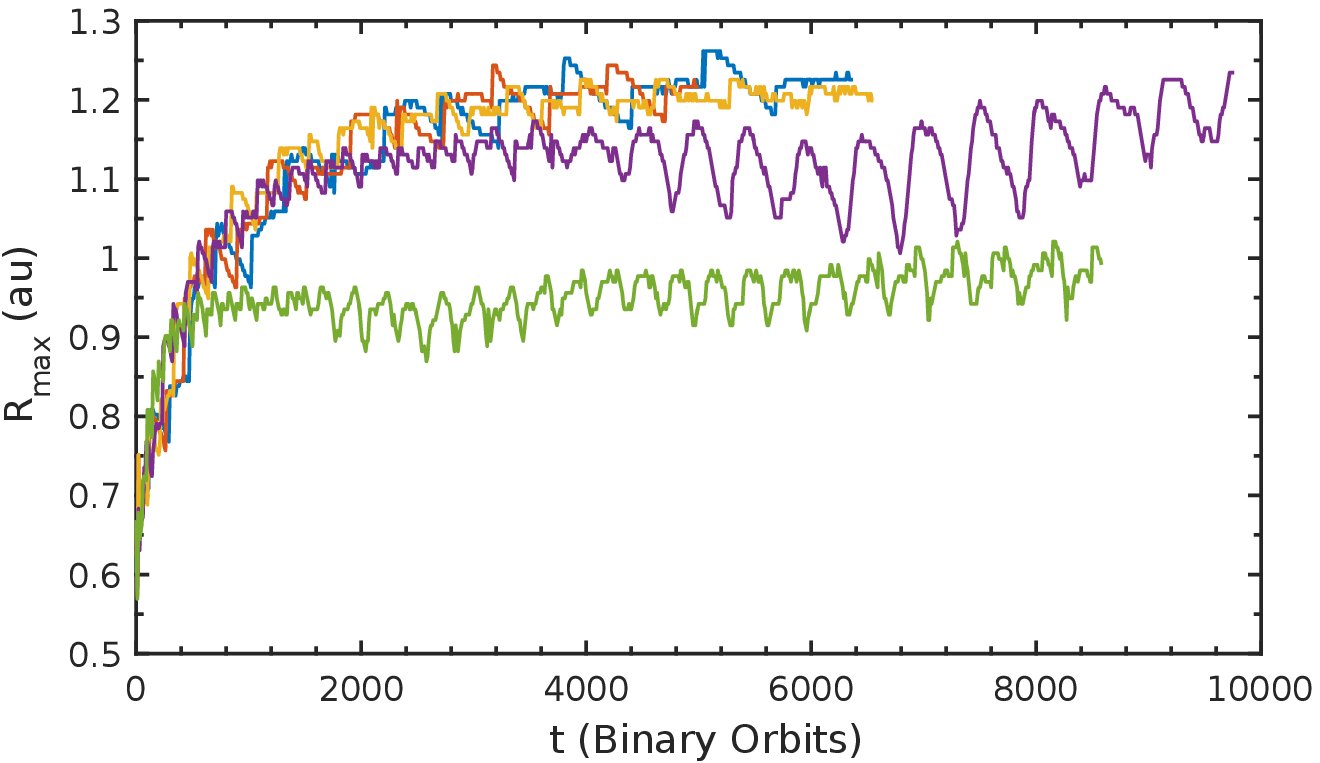}\label{fig:sg_16_rmt}}\\
	\vspace{-13pt}
	\subfloat[]{\includegraphics[width = 0.455\textwidth]{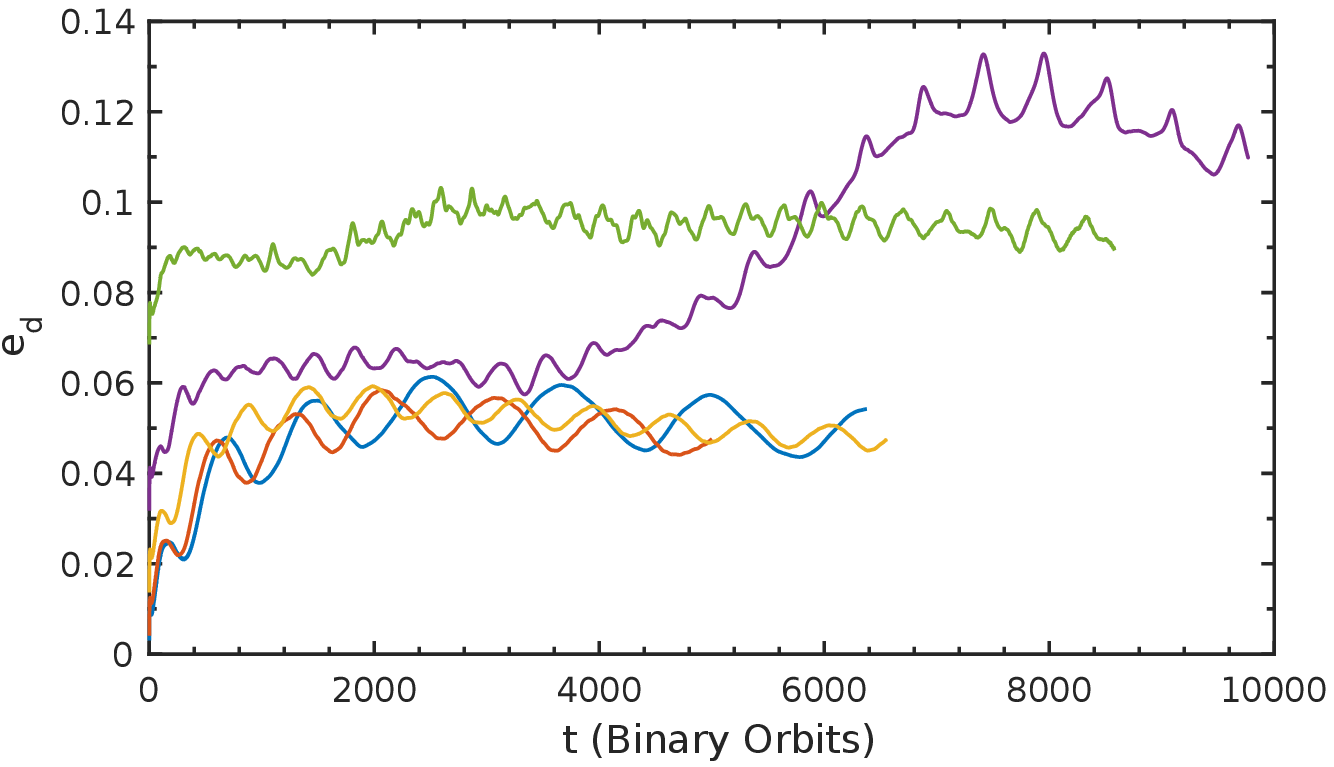}\label{fig:sg_16_edt}}
	\caption{Disc Structure and Dynamics results at pseudo-steady-state, after \Pb{5000}, in the Kepler-16 system.
\textit{a)} Scaled surface density distribution.
\textit{b)} Azimuthally-averaged cell eccentricity.
\textit{c)} Evolution of peak surface density position.
\textit{d)} Evolution of disc-integrated eccentricity.}
	\label{fig:sg_16}
\end{figure}

\subsubsection{Low-Mass Discs}\label{sec:sg_lowm16}

In the Kepler-16 system increasing the disc mass from a canonical \mmsn{1} model to an intermediate disc mass corresponding to \mmsn{5}, makes little difference to the disc evolution or final disc structure. The low-mass discs show no departure from the non-SG $R_\mathrm{in} = 0.09$ au disc seen in the previous section. The pseudo-steady-state cavity size observed agrees well with the PN13 results. A representative surface density map of the low-mass disc results, once they have reached steady-state, can be seen in Fig. \ref{fig:sg_16_1}.

The blue, red and yellow lines in Fig. \ref{fig:sg_16} correspond to the 1, 2, and \mmsn{5} models respectively. In Fig. \ref{fig:sg_16_sd} the surface density distributions have been scaled by their inverse disc mass so their overall shape can be compared. Beyond small changes in the relative peak value, the position and size of the disc cavity is nearly identical, reaching around $1.25$ au at equilibrium. The evolution of the cavity size is similar between these three models -- Fig. \ref{fig:sg_16_rmt}. Increasing the disc mass has no effect on the rate at which the eccentric cavity grows in size as the simulations progress, with all models reaching a pseudo-steady state after $\approx$ \Pb{4000}.

The discs in these models can be separated into three sections in radius: the cavity -- the region where material is evacuated or the density gradient is positive because of the dominant influence of the binary; the inner disc, where the density gradient is negative, but eccentricity is excited; and the outer disc, where eccentricity is negligible, and the disc resembles that of a single-star case (we note that the rise in eccentricity in this outer region as the disc mass increases is largely an artefact of not including the disc self-gravity in the calculation of the eccentricity). In the low-mass regime the cavity is dominated by the binary's influence, but the increasing strength of the disc's self-gravity leads to a slightly more tightly-bound cavity. The self-gravity acts to compact the system, shrinking the characteristic scale slightly compared to the non-SG case. It is noticeable in Fig. \ref{fig:sg_16_e} that the inner region where the eccentricity is excited reduces as disc mass grows. The disc dynamics, traced by the eccentricity gradient, seem to be more sensitive to increasing disc self-gravity than the disc density structure.

Using the last panel in Fig. \ref{fig:sg_16} we note several features. The global disc eccentricity growth rate is slightly faster as the disc mass increases, with all the low-mass discs reaching peak \e{d} ($\approx 0.05\approx h$) at similar times. The oscillation of the disc eccentricity reaches lower amplitudes and shorter periods as the disc mass increases, caused by the circulation of the discs' longitude of pericentre relative to the binary's. The accelerations arising from self-gravity force faster circulation as the disc gravity increases the precession rate, but drive smaller amplitude oscillations in the eccentricity.

\begin{figure}
	\centering
	\includegraphics[width = 0.45\textwidth]{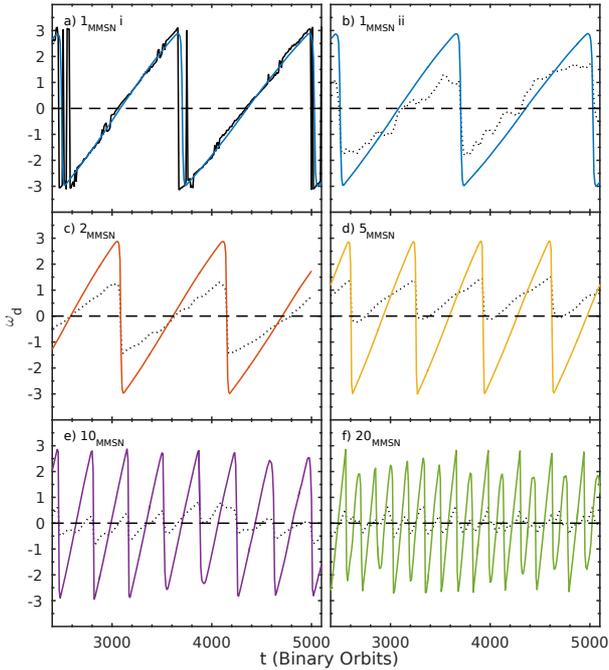}
	\caption{Snapshots of $\omega_\mathrm{d}$ evolution in the Kepler-16 low- and high-mass discs. Panel \textit{a} shows the position angle of the surface density maxima ($+\pi$) in black, overplotted with the local calculation of $\omega_\mathrm{d}$ . Panels \textit{b}--\textit{f} show the local calculation of $\omega_\mathrm{d}$ compared to the global calculation (black-dotted line), in the 1--\mmsn{20} discs.}
	\label{fig:sg_16_wdt}
\end{figure}

Figure \ref{fig:sg_16_wdt} shows the $\omega_\mathrm{d}$ evolution in the Kepler-16 self-gravitating discs. We include a comparison of the local \textit{vs.} global calculation (discussed in Section \ref{sec:varrin_res}) -- as well as comparison with the maximum surface density position angle. Our results agree with those from both PN13 and KH15, where the global calculation gives a near constant value of $\omega_\mathrm{d}$, but the local calculation and the maximum position angle show circulation. In Fig. \ref{fig:sg_16_hight} lines are plotted from the origin, pointing to the local $\omega_\mathrm{d}$ (red) and the maximum position angle + $\pi$ radians (green). The two lines may be difficult to resolve as they are in very close agreement. Comparing the evolution of $\omega_\mathrm{d}$ in the low-mass discs, we observe the same trend as we saw in the oscillatory nature of \e{d}. As the disc mass increases, the periods for both the precession of $\omega_\mathrm{d}$, and the oscillations in \e{d} decrease by nearly a factor of 2, from $\approx 1200$ to \Pb{700}. As the forcing from the binary is not increasing, nor is there any additional component from the pressure forces in the disc, this increase in frequency must be due to the increasing strength of the disc self-gravity. This difference in precession frequency is the clearest impact of self-gravity on disc evolution for low-mass discs in Kepler-16 type systems.

\subsubsection{High-Mass Discs}\label{sec:sg_highm16}

\begin{figure}
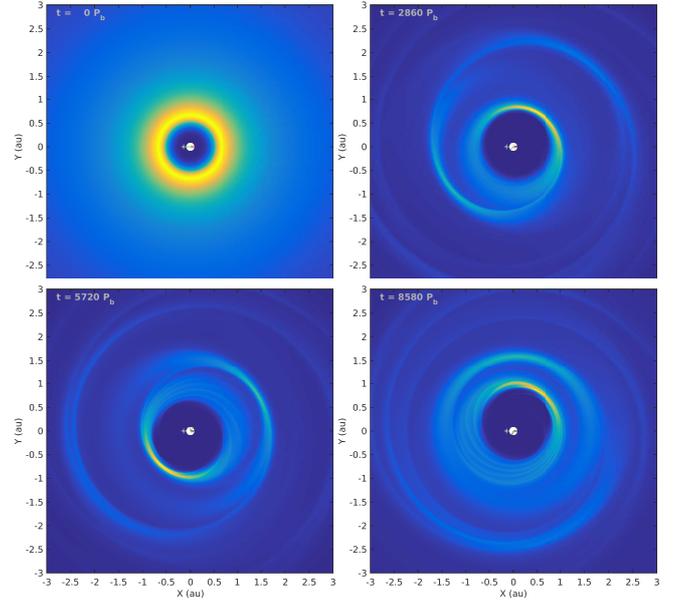

	\centering
	\subfloat{\includegraphics[width = 0.24\textwidth]{./figures/surf/16/SG20_sd0.eps}\label{fig:sg_16_sd20_1}}
	\subfloat{\includegraphics[width = 0.24\textwidth]{./figures/surf/16/SG20_sd286.eps}\label{fig:sg_16_sd20_2}}\\ \vspace{-25pt}
	\subfloat{\includegraphics[width = 0.24\textwidth]{./figures/surf/16/SG20_sd572.eps}\label{fig:sg_16_sd20_3}}
	\subfloat{\includegraphics[width = 0.24\textwidth]{./figures/surf/16/SG20_sd858.eps}\label{fig:sg_16_sd20_4}}
	\caption{Surface density plots of showing evolution of highly self-gravitating discs in the Kepler-16 \mmsn{20} system. The birth and evolution of a series of nested eccentric rings and spirals can be seen.}
	\label{fig:sg_16_20_grid}
\end{figure}

Whilst the low-mass discs showed little variation across the mass range, and from previously obtained results, it is clear from Fig. \ref{fig:sg_16} that increasing the disc mass to \mmsn{10}, and beyond, leads to dramatic modifications in disc structure, dynamics and evolution compared to the low-mass discs. Some of these changes are shared across high-mass disc models in all the Kepler systems, suggesting that they are caused predominantly by the disc self-gravity, with the binary properties having little impact. We will describe these features in turn for each Kepler system. 

Doubling the disc mass from \mmsn{5} to \mmsn{10} in the Kepler-16 system, brings us into a regime where the strength of the disc's self-gravity can start to significantly alter the disc structure seen in the low-mass discs. Referring to the purple line in Fig. \ref{fig:sg_16_sd} we see an enhancement in the relative peak surface density, as well as a decrease in $R_\mathrm{max}$. The surface density falls away sharply exterior to this peak, reaching a minimum at $\approx 1.65$ au, only to rise to a second smaller peak at 2 au. Beyond this point the disc relaxes back to the background profile seen in the low-mass discs. The disc structure is modified even further in the \mmsn{20} disc (green line in Fig. \ref{fig:sg_16_sd}), with the relative peak surface density obtaining a value twice that seen in the low-mass discs, and moving interior to 1 au. In the exterior of the disc, a series of two further surface density maxima can be seen, at 1.6 and 2.5 au. These objects are maintained by disc self-gravity, overcoming the opposing influences of the gas's pressure and viscosity. The variation in surface density caused by these features is clear when examining the 2D surface density plots of the disc. Figure \ref{fig:sg_16_20_grid} shows the evolution of the disc in the \mmsn{20} model. The maxima observed in the surface density profiles are associated with a series of nested eccentric features which at times resemble eccentric rings, similar in form to the material bounding the inner cavity, and at other times resemble a large scale $m=1$ spiral wave. It is also clear in these plots that the inner cavity, although eccentric, has not expanded significantly in size since the disc was initialised. In this model the strength of the self-gravity in the disc results in a significantly more compact inner disc cavity region.

These structures can also be seen in the discs' eccentricity profiles (Fig \ref{fig:sg_16_e}). The \mmsn{10}and \mmsn{20} models have similar eccentricity values in the binary-dominated cavity -- only in a smaller region -- with eccentricity minima and maxima in the outer disc. The eccentricity in the outer disc is excited above the non-negligible values common to the low-mass discs, with \e{c}$ \approx 0.1$. The peaks in surface density seen in Fig. \ref{fig:sg_16_sd} correspond to minima in the \e{c} profile, with maxima on either side. These surface density peaks are obviously associated with further eccentric structures in the disc, exterior to the eccentric inner cavity.

The extra sources of \e{c} are evident as raised values in the evolution and final value of global \e{d}. Figure \ref{fig:sg_16_edt} shows more complex behaviour for the evolution of \e{d} than in the low-mass cases. At early times the \mmsn{10} model shows oscillatory behaviour around \e{d} similar to the low-mass discs. The slightly higher value is due to the raised \e{c} in the outer disc. At $t = $ \Pb{3000}, once the inner cavity has fully developed, \e{d} starts to increase, reaching a peak value around 0.13, and saturating at a final value close to 0.12. The growth of \e{d} is due to the creation of the additional eccentric features seen in \Cref{fig:sg_16_sd,fig:sg_16_e,fig:sg_16_20_grid}. The \mmsn{20} model starts with initial \e{d} $\approx 0.08$, due to the contribution of self-gravity on the dynamics that is not accounted for in the calculation of disc eccentricity. The disc eccentricity is in a pseudo-steady state until $t = $ \Pb{1000} when \e{d} increases for \Pb{2000} until it saturates at 0.1. Strong self-gravity in the \mmsn{20} disc explains the earlier onset of the second phase of \e{d} growth. The overall smaller value of \e{d} versus that in the \mmsn{10} model can be explained by the fact that the inner cavity has a much smaller radial extent -- the material beyond 1 au, with \e{c}$\approx0.1$ makes up a large proportion of the disc, containing much of the mass in the disc, so dominates \e{d}. 

The very short-term oscillations in \e{d} are caused by the circulation of the disc around the binary's fixed longitude of pericentre. The period of these oscillations is the same as the period taken for the inner cavity to circulate a full $2\pi$ radians. With this evidence we would predict that the high-mass discs follow the trend seen in the low-mass discs -- with the period decreasing as disc mass increases. Examining panels \textit{e} and \textit{f} in Fig. \ref{fig:sg_16_wdt}, we see this is basically true. These more massive discs don't show perfect circulation, but instead show evidence of libration. Remember that the local calculation of $\omega_\mathrm{d}$ only includes material upto $\Sigma_\mathrm{max}$ in the azimuthally-averaged profile. Therefore the gas associated with the additional structures seen in these discs is not directly affecting $\omega_\mathrm{d}$. It is possible for these features to interact with the cavity-bounding material -- through self-gravity -- disrupting its circulation. This, as well as the non-constant extent of libration, suggests that the features in the outer disc have their own associated precession period, independent of the inner disc, which we discuss later in the paper.

\subsection{Kepler-34}\label{sec:sg_34}

\begin{figure}
	\centering
	\subfloat[]{\includegraphics[width = 0.42\textwidth]{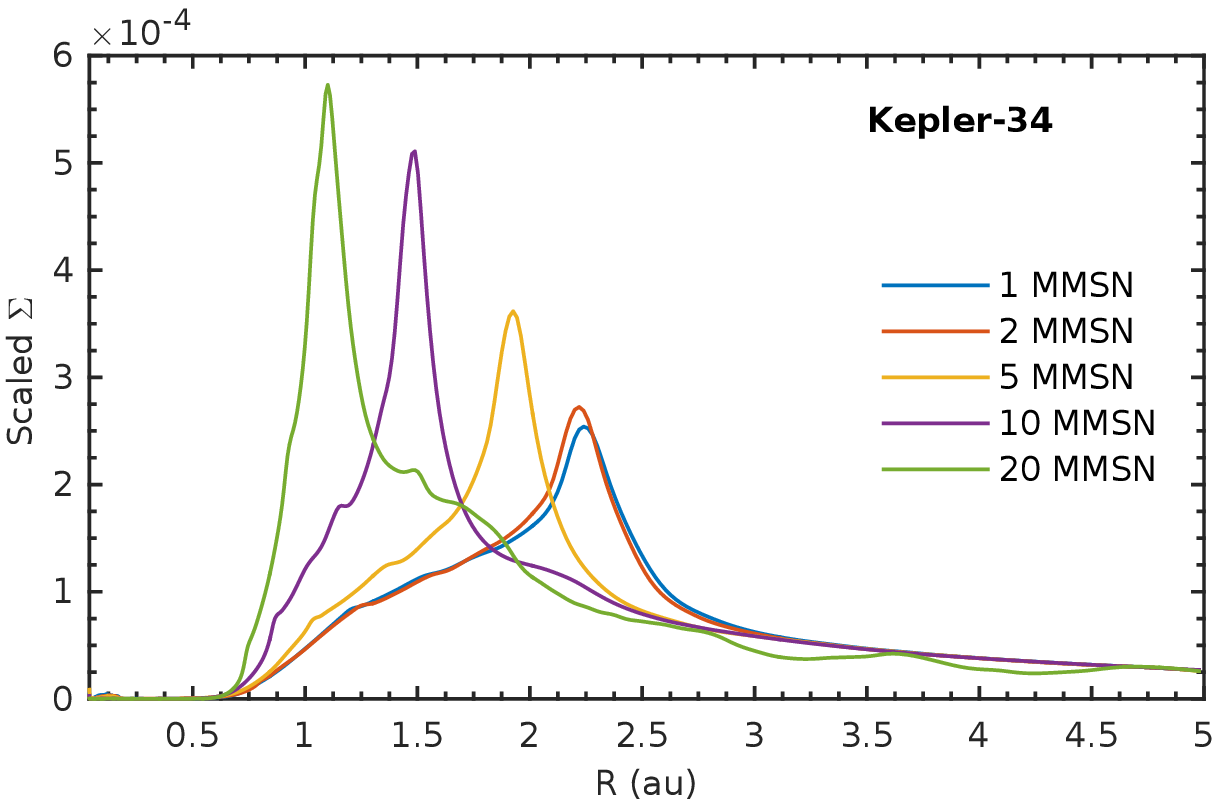}\label{fig:sg_34_sd}}\\
	\vspace{-13pt}
	\subfloat[]{\includegraphics[width = 0.44\textwidth]{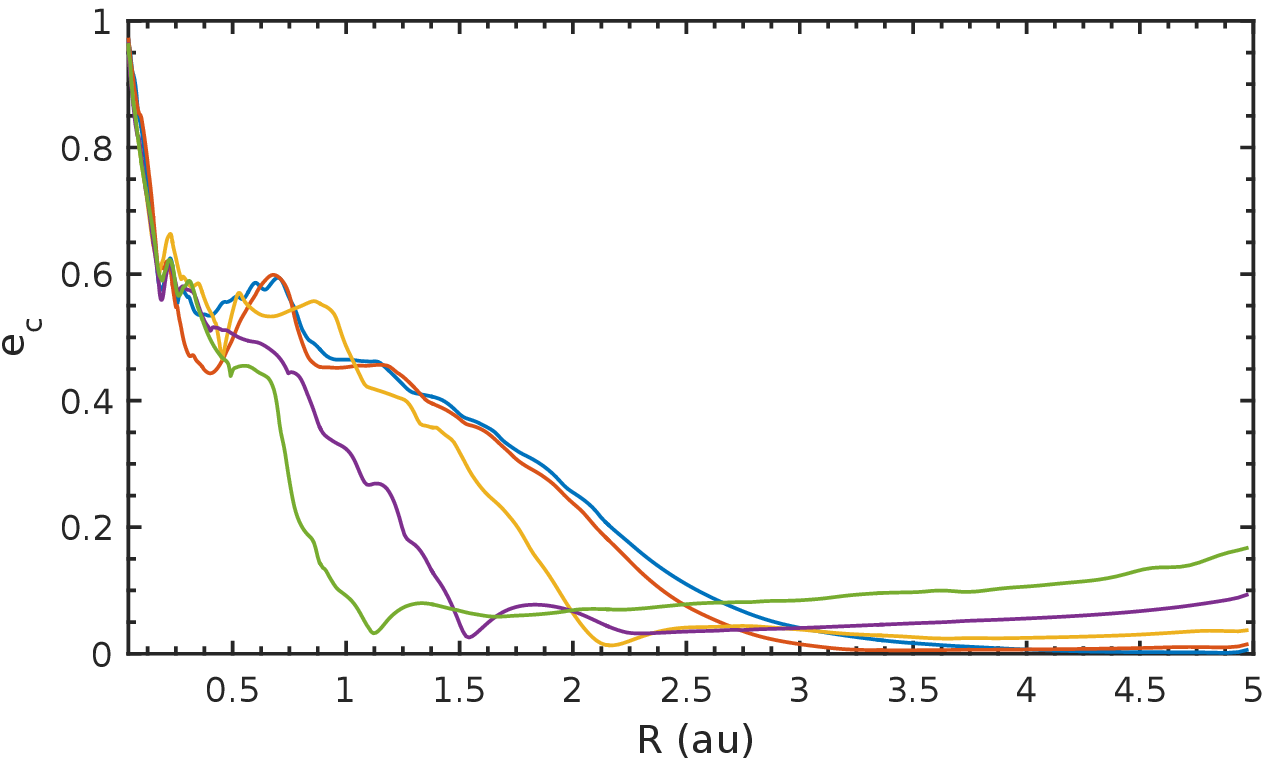}\label{fig:sg_34_e}}\\
	\vspace{-13pt}
	\subfloat[]{\includegraphics[width = 0.46\textwidth]{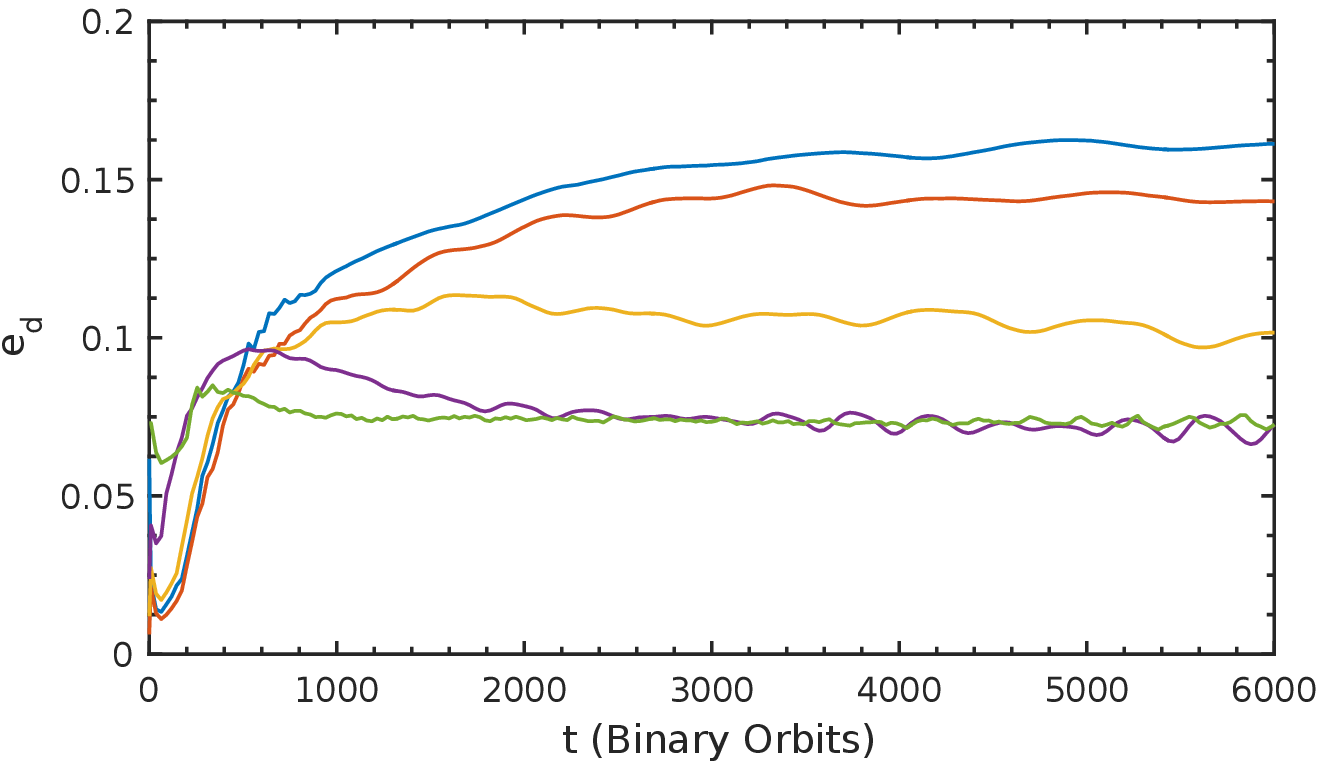}\label{fig:sg_34_edt}}
	\caption{Disc Structure and Dynamics results at pseudo-steady-state, after \Pb{6000}, in the Kepler-34 system. \textit{a)} Surface Density distribution. \textit{b)} Azimuthally-averaged cell eccentricity. \textit{c)} Evolution of disc-integrated eccentricity.}
	\label{fig:sg_34}
\end{figure}

\subsubsection{Low-Mass Discs}\label{sec:sg_lowm34}

The low-mass discs in the Kepler-34 analogue systems show the most variation across all the low-mass models in the three systems investigated here. The \mmsn{5} disc around this highly eccentric binary departs from the behaviour observed in the \mmsn{1} and \mmsn{2} discs. As mentioned previously, the Kepler-34 system produces a very eccentric disc cavity, extending out to $\approx2.3$ au -- slightly larger than the Kepler-34 results shown in PN13, KH15 and \citet{Lines2015}. In Fig. \ref{fig:sg_34_sd} the yellow line, corresponding to the \mmsn{5} model, shows a smaller cavity size ($R_\mathrm{max} = 2.0$) and higher relative $\Sigma_\mathrm{max}$, by a factor of 1.25. This is behaviour we are observing for the first time, i.e. this behaviour was not seen in the Kepler-34 self-gravitating discs in \citet{Lines2015}. The eccentricity distribution in the \mmsn{5} case shows material on highly excited orbits in the cavity itself, with a fast drop-off in \e{c} as we move outwards in the disc. The disc self-gravity leads to a more radially confined eccentric cavity. The influence of the binary beyond $\approx 1.5$ au appears to be small. 

The \mmsn{1} and \mmsn{2} disc structures are very similar, with the doubling of mass having little effect on the structure or evolution. The final values of \e{d} in the \mmsn{1} and \mmsn{2} models reach values of $\approx 0.15$, higher than the value found in PN13 for similar disc parameters. These values are predictably higher than the Kepler-16 models of the same mass, due to Kepler-34's higher binary eccentricity. The \mmsn{5} disc reaches \e{d} = 0.1 at equilibrium, due to the smaller values of \e{c} between 2 and 3 au, and the smaller radial extent of the cavity. Increasing the disc mass, and consequently the strength of self-gravity, acts to decrease the radial scale of the disc system, as observed for the Kepler-16 analogue system discussed above. The general results from these low-mass discs further corroborate previous findings suggesting that the final, saturated value of \e{d} -- as well as the cavity size -- are strongly dependent on binary eccentricity, and not mass ratio. 

The discrepancy seen in \e{d} in Fig. \ref{fig:sg_34_edt} between the \mmsn{1} and \mmsn{2} models suggests that small changes in the calculated \e{c} distribution in the outer disc can make a significant difference to the global value of disc eccentricity, since this is where most of the mass is. Perhaps, to discount the region of the disc where little interaction with the binary occurs, a local calculation of \e{d} may highlight the evolution of the eccentric inner disc more accurately. There is little evidence of oscillations in the evolution of \e{d} for any of these low-mass discs, except for a very long period, low amplitude trend in the \mmsn{5} case.

\subsubsection{High-Mass Discs}\label{sec:sg_highm34}

\begin{figure}
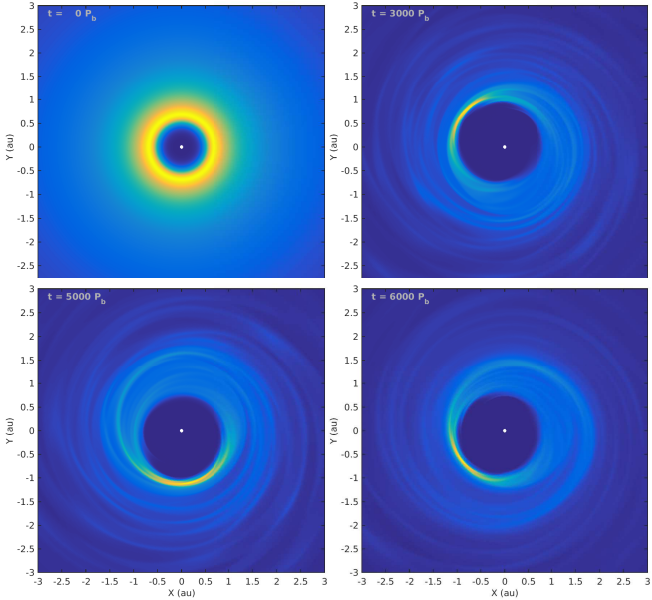

	\centering
	\subfloat{\includegraphics[width = 0.24\textwidth]{./figures/surf/34/SG20_sd0.eps}\label{fig:sg_34_sd20_1}}
	\subfloat{\includegraphics[width = 0.24\textwidth]{./figures/surf/34/SG20_sd300.eps}\label{fig:sg_34_sd20_2}}\\ \vspace{-25pt}
	\subfloat{\includegraphics[width = 0.24\textwidth]{./figures/surf/34/SG20_sd500.eps}\label{fig:sg_34_sd20_3}}
	\subfloat{\includegraphics[width = 0.24\textwidth]{./figures/surf/34/SG20_sd600.eps}\label{fig:sg_34_sd20_4}}
	\caption{Surface density plots of showing evolution of highly self-gravitating discs in the Kepler-34 \mmsn{20} system. Compared to the same model in the Kepler-16 system, the additional features seen here are not as clearly defined.}
	\label{fig:sg_34_20_grid}
\end{figure}

The Kepler-34 high-mass discs continue the trend seen in the low-mass discs. As we move to the \mmsn{10} model, the influence of the self-gravity causes $R_\mathrm{max}$ to reduce quite dramatically to 1.5 au -- significantly smaller than the non-SG results from PN13 and \citet{Lines2015} -- and a value of 1.1 au in the \mmsn{20} disc. The surface density distributions (Fig. \ref{fig:sg_34_sd}) are much smoother than those seen in the Kepler-16 high-mass discs (Fig. \ref{fig:sg_34_sd}). Whilst there are no sharp peaks, there is evidence of less dramatic bumps in the profiles; the same features seen in Kepler-16 are spread out over a wider radial area.

In Fig. \ref{fig:sg_34_20_grid} we present 2D surface density snapshots for the evolution of the Kepler-34 \mmsn{20} disc, in the same format as the previous grid for Kepler-16. Significant modification to the outer disc beyond the eccentric inner cavity can be seen. The washed-out bumps seen in Fig. \ref{fig:sg_34_sd} are also clear in this figure. The eccentric rings described previously for the Kepler-16 10 and \mmsn{20} models look more like a series of interacting spiral-arms in the Kepler-34 system, but nonetheless the development of an eccentric ring feature with an $m=1$ non-axisymmetric surface density enhancement can be observed.

\begin{figure}
	\centering
	\includegraphics[width = 0.45\textwidth]{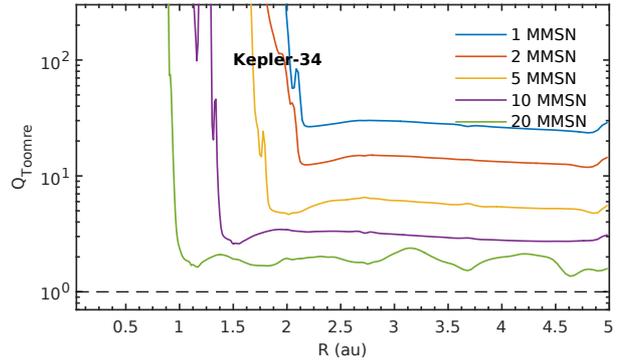}
	\caption{Radial profiles of the Toomre parameter corresponding to evolved disc conditions in the Kepler-34 system, with masses corresponding to 1 -- \mmsn{20}. Whilst the low-mass discs retain $Q$ values $> 1$, the high-mass discs show regions where $Q \approx 1$.}
	\label{fig:toomreq_evol}
\end{figure}

An examination of the Toomre parameter profiles (Fig. \ref{fig:toomreq_evol}) in these evolved massive discs reveals values very close to 1 in the high density features. Unsurprisingly the minima in $Q$ correspond to surface density maxima -- the high densities seen at the apocentres of the eccentric rings in the disc give rise to small values of the Toomre parameter. Although the Toomre $Q$ value approaches unity, it is worth noting that we see no evidence of disc fragmentation in these massive discs. We suspect that further increases in disc mass may, however, lead to disc fragmentation, as we do see signs of low amplitude spiral wave structure developing early in the simulations associated with the disc response to self-gravity for the most massive discs.

The evolution of \e{d} is also different in this system (see Fig. \ref{fig:sg_34_edt}) compared to the same discs in the Kepler-16 system. Both the \mmsn{10} and \mmsn{20} discs go through a short period of growth -- although this is very short in the most massive disc -- followed by \Pb{1000} of slow decline. Both discs then settle into a final value of \e{d} = 0.07 -- smaller than the low-mass discs. Examining the radial profiles of \e{c} in Fig. \ref{fig:sg_34_e}, we see the shrinking of the eccentric cavity caused by the compaction associated with the more massive discs.

\subsection{Kepler-35}\label{sec:sg_35}

\begin{figure}
	\centering
	\subfloat[]{\includegraphics[width = 0.44\textwidth]{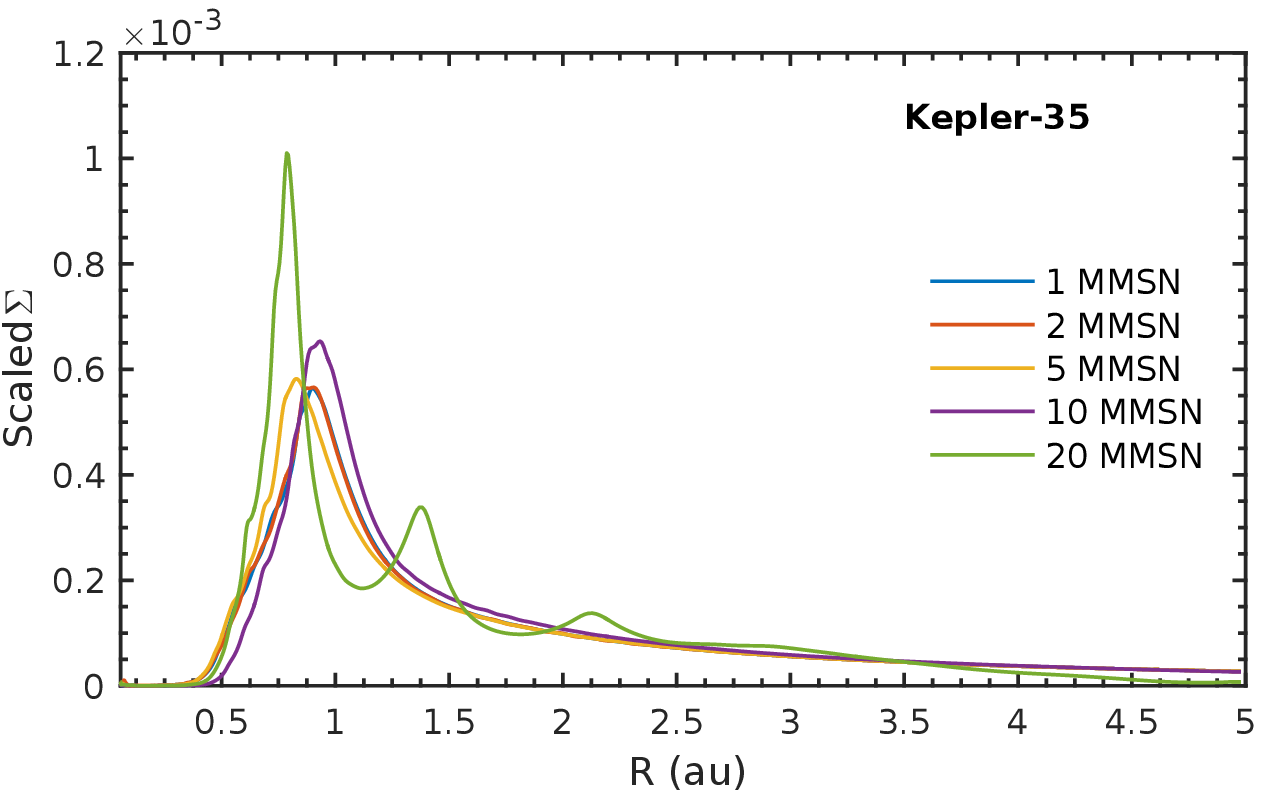}\label{fig:sg_35_sd}}\\
	\vspace{-13pt}
	\subfloat[]{\includegraphics[width = 0.44\textwidth]{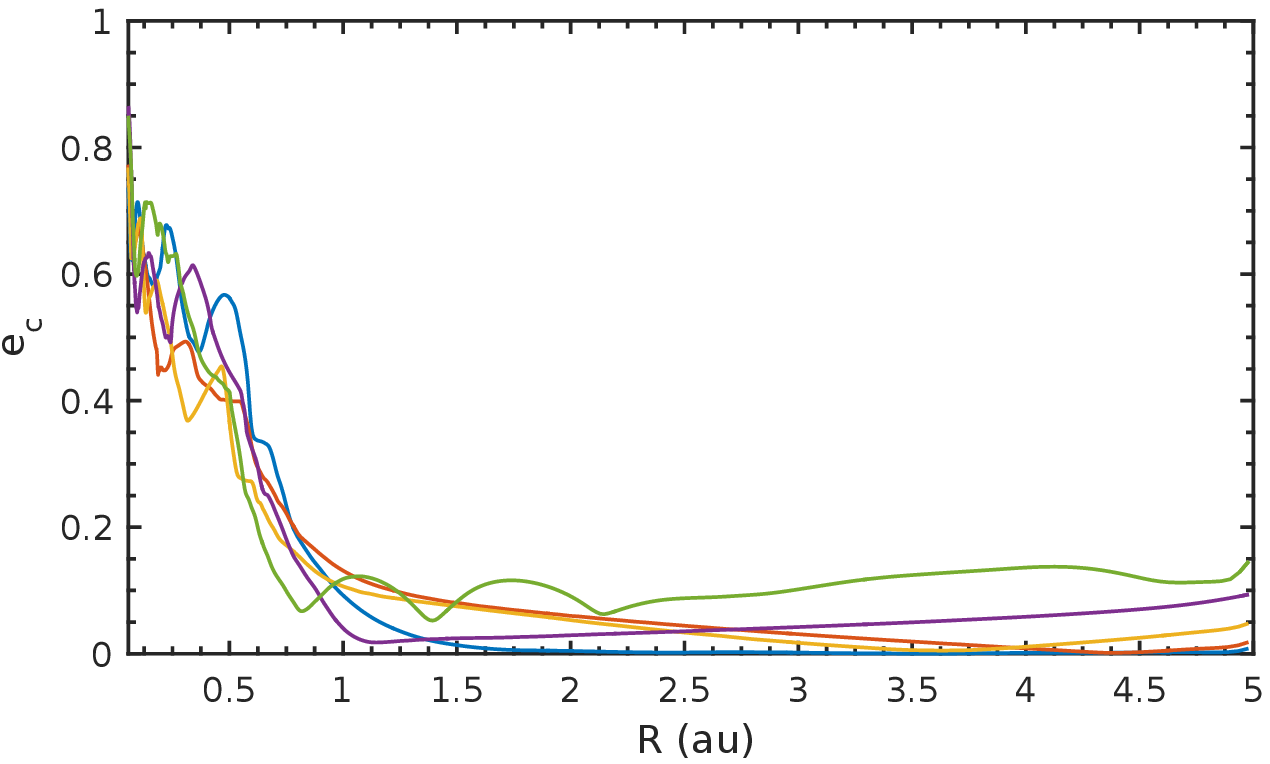}\label{fig:sg_35_e}}\\
	\vspace{-13pt}
	\subfloat[]{\includegraphics[width = 0.46\textwidth]{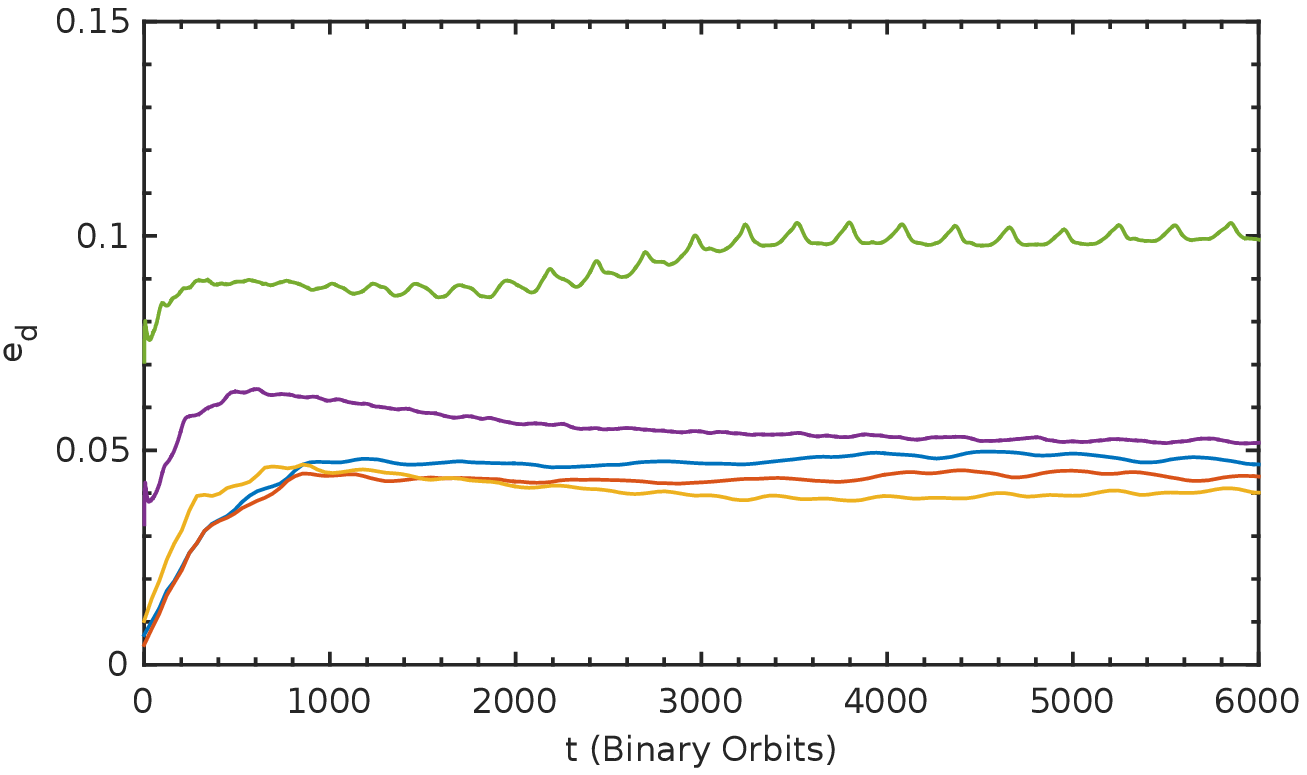}\label{fig:sg_35_edt}}
	\caption{Disc Structure and Dynamics results at pseudo-steady-state, after \Pb{6000}, in the Kepler-35 system.
	\textit{a)} Surface Density distribution.
	\textit{b)} Azimuthally-averaged cell eccentricity.
	\textit{c)} Evolution of disc-integrated eccentricity.}
	\label{fig:sg_35}
\end{figure}

\subsubsection{Low-Mass Discs}\label{sec:sg_lowm35}

Kepler-35 AB has a similar eccentricity to Kepler-16, but a mass ratio comparable to Kepler-34, so we would expect disc structure and evolution to be close to those seen in the Kepler-16 discs. This seems to be the case when we examine the surface density and \e{c} profiles in \Cref{fig:sg_35_sd,fig:sg_35_e}. The smaller \e{b} and $a_\mathrm{b}$ lead to a smaller cavity, with $\Sigma_\mathrm{max}$ lying just interior to 1 au. The \e{c} profile mirrors this structure, with the eccentric inner region confined within 1 au, and negligible differences in eccentricity gradient. The surface density and \e{c} profiles are in good agreement with those seen for the results from PN13 with similar disc parameters. The first two panels in Fig. \ref{fig:sg_35_grid} show surface density plots in the \mmsn{1} and \mmsn{5} models at steady-state. These plots highlight the similarity in cavity size and disc structure found in the low-mass models in low-eccentricity binary systems.

The evolution of \e{d} shows no evidence of periodic oscillations, although the sampling frequency could hide finer oscillations. This lack of oscillatory behaviour seems to be the norm in the close-to-unity mass ratio binaries (i.e. Kepler-34 and -35). This may be explained by the symmetry of the binary orbit -- the central stars are similar distances from the binary's center of mass at both closest and furthest approach. The inner eccentric disc does not get as disturbed periodically as much as in the lower mass ratio Kepler-16 binary, where circulation of the eccentric disc causes periodic apsidal alignment and misalignment between the binary and disc. Finally, examining Fig. \ref{fig:sg_35_edt}, we see that these low-mass discs reach similar \e{d} values as the Kepler-16 models, as well as those seen in PN13.

\subsubsection{High-Mass Discs}\label{sec:sg_highm35}

The last set of simulations from our systematic investigation into the role of disc mass and self-gravity in the Kepler circumbinary systems are the high-mass discs in Kepler-35. Interestingly the peak surface density position in these discs is similar to those found in the low-mass regime -- a finding not duplicated in the other systems -- with all the peaks lying close to 1 au. Nonetheless, the trend of having a smaller more compact inner cavity for a more massive disc is reproduced in this system, albeit less dramatically than for Kepler-16 and -34.

\begin{figure}
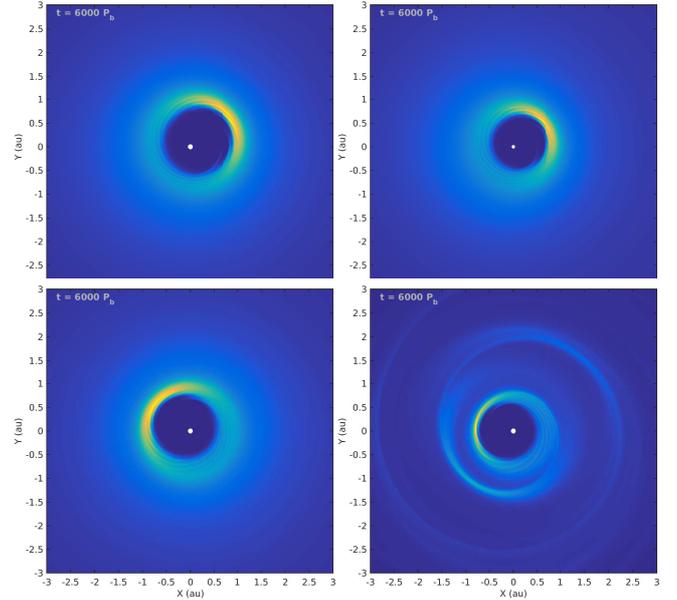

	\centering
	\subfloat{\includegraphics[width = 0.24\textwidth]{./figures/surf/35/SG1_sd600.eps}\label{fig:sg_35_sd1}}
	\subfloat{\includegraphics[width = 0.24\textwidth]{./figures/surf/35/SG5_sd600.eps}\label{fig:sg_35_sd5}}\\ \vspace{-25pt}
	\subfloat{\includegraphics[width = 0.24\textwidth]{./figures/surf/35/SG10_sd600.eps}\label{fig:sg_35_sd10}}
	\subfloat{\includegraphics[width = 0.24\textwidth]{./figures/surf/35/SG20_sd600.eps}\label{fig:sg_35_sd20}}
	\caption{Surface density plots of self-gravitating discs in the Kepler-35 system. A tight inner cavity can be seen in the low-mass \mmsn{1} (top-left) and \mmsn{5} (top-right) discs, as well as the high-mass \mmsn{10} (bottom-left) model. The nested eccentric rings seen in the Kepler-16 \mmsn{10} and \mmsn{20}, and Kepler-34 \mmsn{20} models can be seen in the highest mass, \mmsn{20} model in this system.}
	\label{fig:sg_35_grid}
\end{figure}

Further eccentric rings beyond 1 au are not excited in the \mmsn{10} model as \Cref{fig:sg_35_sd,fig:sg_35_sd10} shows, although we cannot discount the possibility that they may arise on longer time scales than we have simulated for this system. There is also no evidence of further excited eccentric features in the \e{c} profile found in Fig. \ref{fig:sg_35_e}, or in the evolution of \e{d} (Fig. \ref{fig:sg_35_edt}). 

Examining the most massive \mmsn{20} disc, we see the appearance of the additional well-defined freely precessing eccentric ring features also found in Kepler-16. The self-gravity in this disc is strong enough for two further eccentric rings to be excited exterior to the inner cavity. This process occurs around \Pb{2000} in the disc's evolution, as this is when we observe the \e{d} of the system start to increase. After a further \Pb{1000} the disc has reached a pseudo-steady equilibrium \e{d} $ = 0.1$. This evolution and final value is similar to Kepler-16, despite this model having a much smaller eccentric inner cavity. 

The respective lack, and presence, of eccentric features in the \mmsn{10}, and \mmsn{20}, models is shown in the last two panels of Fig. \ref{fig:sg_35_grid}. The similarity of the \mmsn{10} case to the 1--\mmsn{5} models is clear, whereas the most massive disc shows the eccentric features found to be a common feature across these self-gravitating discs. 

\subsection{Origin of the Additional Eccentric Features in High-Mass Discs}\label{sec:sg_origin}

\begin{figure}
	\includegraphics[width = 0.44\textwidth]{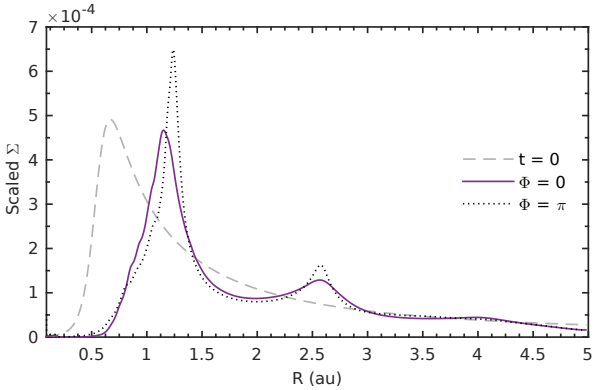}
	\caption{Interaction between two precessing eccentric rings through self-gravity. The purple profile is when the Longitudes of Pericentre of each ring are aligned, black when they are anti-aligned, or separated by $\pi$ radians.}
	\label{fig:sg_16_10}
\end{figure}

In this section we investigate the origin of the additional eccentric rings in our most massive circumbinary discs. We consider two contending ideas: (i) That secular gravitational interaction between the eccentric inner disc at the tidally cleared cavity edge and the outer disc leads to the excitation of an additional eccentric ring if the local precession frequencies in the disc are able to match (i.e. the additional eccentric feature results from a secular resonance in the self-gravitating disc); (ii) That the binary potential itself is responsible for the growth of the additional eccentric features, possibly through non-linear mode coupling similar to that occurring at the 3:1 resonance as examined by \cite{Papaloizou2001} and \cite{Pierens2013} in non-self-gravitating discs, but occurring over larger radial length scales in the self-gravitating discs, possibly because changes to the epicyclic frequencies shift the resonances outwards.

When considering scenario (i), we undertook an analysis of the rings' precession. By finding the surface density maxima in the azimuthally-averaged surface density profile, and isolating material in narrowly-defined annuli around these radial positions, we were able to obtain rough estimates of the individual precession periods of each eccentric ring. In the Kepler-16 system, \mmsn{10} disc, the inner eccentric cavity precesses with a period $ = $ \Pb{420}. The first outer eccentric ring has a precession period almost 6 times greater, at \Pb{2440}. The second (and last) outer eccentric feature in this disc also has this same larger precession period, and for each system where two additional eccentric features arise we find that they precess at the same rate as each other. Undertaking this analysis for the more massive \mmsn{20} discs in the Kepler-16 and -35 systems, which also show clearly defined outer eccentric rings, we see a similar pattern of behaviour. As we have already observed, the inner eccentric feature precesses rapidly in these discs -- \Pb{170} and \Pb{180} in the Kepler-16 and -35 systems respectively. The first and second outer eccentric rings precess with a period almost exactly 2 times greater than the inner feature in the Kepler-16 system, with both features having a precession period $ \approx$ \Pb{340}. The outer rings' precession period in the Kepler-35 system lies just outside this commensurability at \Pb{440}.

As these outer rings and the inner eccentric cavity precess with their own frequency, they interact with each other. This interaction can be seen in Fig. \ref{fig:sg_16_10} -- when the rings are aligned in the same quadrant of the disc they act to diminish each other, resulting in reduced surface density maxima, and vice-versa when they are anti-aligned (or in opposite disc quadrants). The interaction between these freely precessing rings also seems to interfere with the circulation of the inner cavity, see panels \textit{e} and \textit{f} of Fig. \ref{fig:sg_16_wdt}. The disparity between the precession frequencies of the various eccentric modes in the discs clearly casts doubt on the idea that the additional eccentric features arise because of secular forcing by the inner eccentric disc, although there does seem to be a commensurate relation between the precession periods of the two outer additional eccentric features in discs where two of these arise.

\begin{figure*}
	\includegraphics[width = 0.94\textwidth]{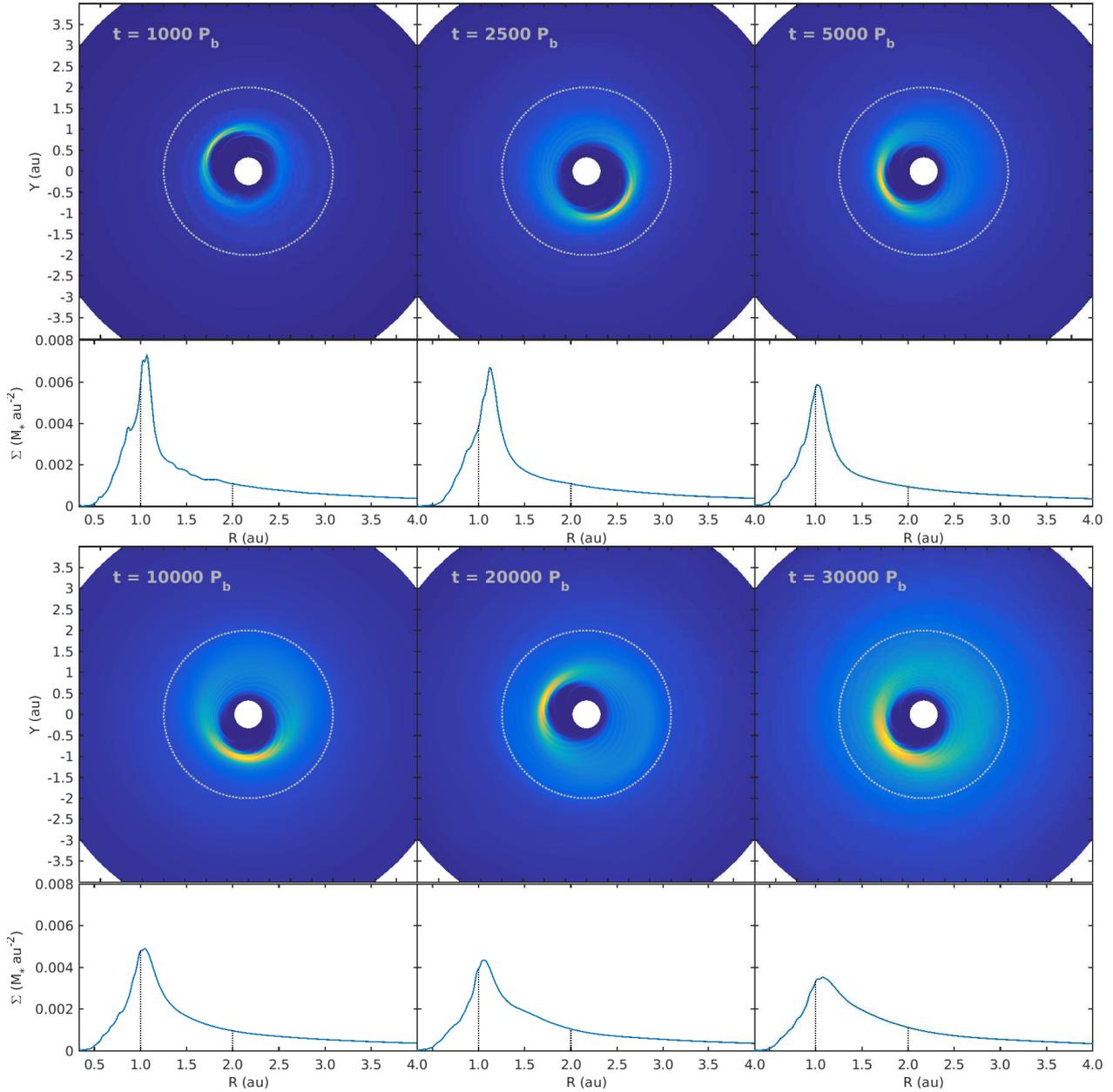}
	\caption{Snapshots of surface density plots and corresponding azimuthally-averaged surface density profiles for the Kepler 16 \mmsn{10}, binary-to-single potential transition (case 1). The grey and black dashed lines show the inner and outer limits of the potential transition at 1 and 2 au. No additional features are seen in the outer disc beyond the central cavity.}
	\label{fig:sg_pt_grid}
\end{figure*}

\begin{figure*}
	\includegraphics[width = 0.94\textwidth]{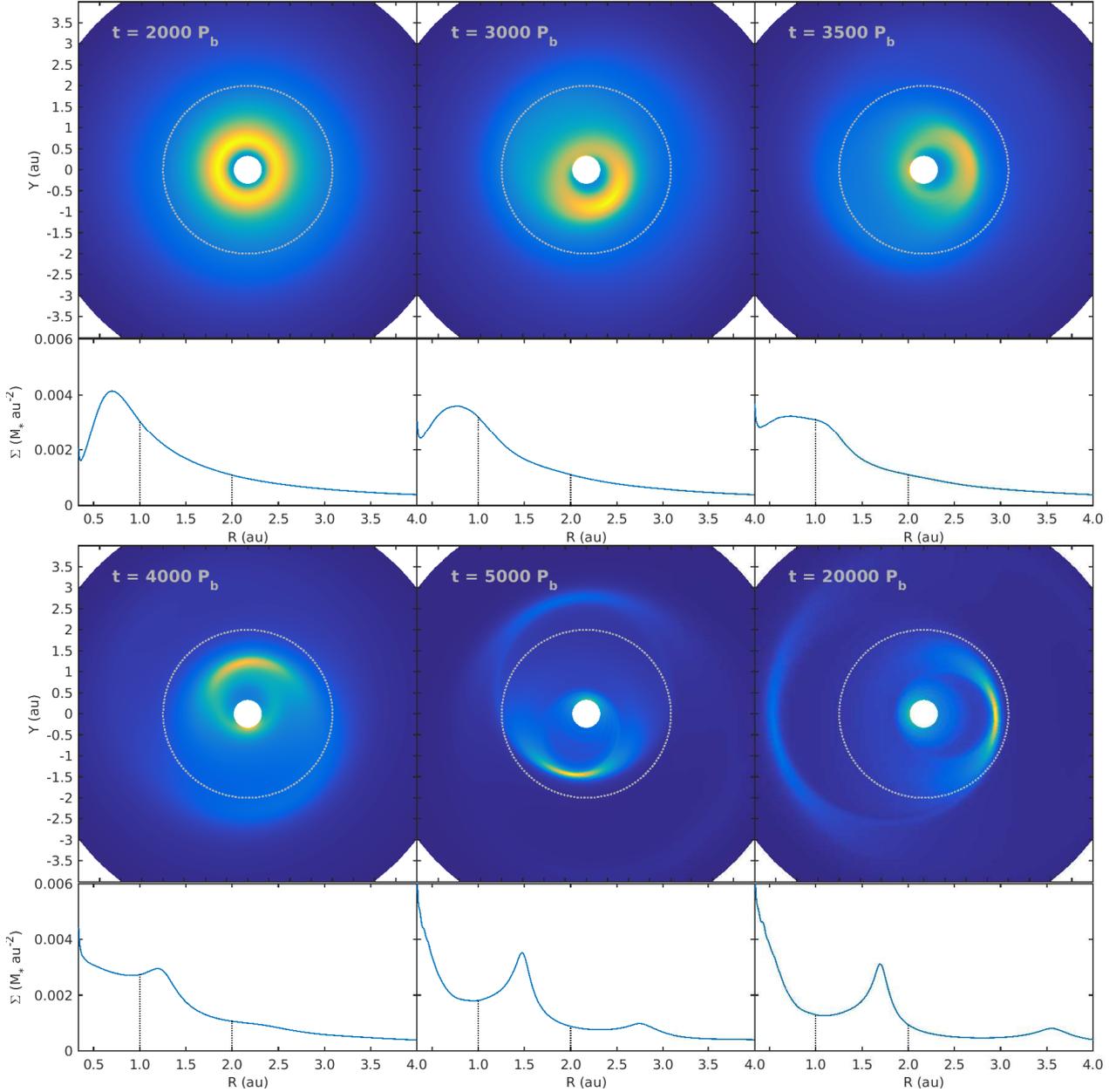}
	\caption{Snapshots of surface density plots and corresponding azimuthally-averaged surface density profiles for the Kepler 16 \mmsn{10}, single-to-binary potential transition (case 2). The grey and black dashed lines show the inner and outer limits of the potential transition at 1 and 2 au. The birth, and growth, of an eccentric feature in the outer binary-dominated region of the disc can be seen.}
	\label{fig:sg_rpt_grid}
\end{figure*}

The second approach to explain these intriguing feature focused on the explicit interaction between the binary and disc. As discussed earlier in the paper, the eccentricity of the inner cavity is driven by non-linear mode coupling between an initial $m=1$ eccentric mode present in the disc and the $m=1$ component of the binary potential, leading to a forcing term in the disc that excites an $m=2$ wave at the 3:1 Lindblad resonance whose outward propagation removes angular momentum from the disc material there. \cite{Pierens2013} also indicated that for an eccentric binary, additional higher order binary potential components may also play a role. The question is whether or not the additional eccentric features also require direct coupling between the disc and the binary potential.

In the additional set of simulations we first use a disc set-up identical to those used in the Kepler-16 \mmsn{10} SG runs presented above, except with a larger inner boundary radius to speed up the computations -- as we are more interested with behaviour in the outer disc, and not an accurate description of the inner disc. The only other difference is that we modify the potential created by the central stars which the disc elements see at different radial locations throughout the disc. In the first case we transition from an inner region where the disc sees a potential created by the two binary stars, to an outer region where the disc sees a potential created by a central single star with the combined mass of the binary, with the cross-over occurring through a transition region of finite width. A second case, where we transition from a single- to binary-dominated potential moving out in the disc, was also computed.

We use a simple switch function from \citet{McNeil2009} of the form, $ f(x) = 3x^2 - 2x^3 $ which transitions smoothly and continuously from $f(x = 0) = 0$ to $f(x = 1) = 1$. We set the scaling factor $x = \frac{R - R_0}{R_1 - R_0}$, where $R$ is the radial location in the disc, and $R_0$ and $R_1$ are the inner and outer limits of the transition region respectively. To ensure we have an outer region sufficiently isolated from the influence of the binary potential we set $R_0 = 1$ and $R_1 = 2$ au.

We define two weighting coefficients which take the following values throughout the disc:
\begin{align}
\begin{split}
	W_\mathrm{binary}(R) &= 
	\begin{cases}
		1 \ \ \ \ \ \ \ \ \ \ \ \ \text{ for}\ R < R_{0} \\
		0 \ \ \ \ \ \ \ \ \ \ \ \ \text{ for}\ R > R_{1} \\
		1 - f(x) \ \text{ for}\ R_{0} \leqslant R \leqslant R_{1},
	\end{cases} \\
	W_\mathrm{single}(R) &= 
	\begin{cases}
		0 \ \ \ \ \ \ \ \ \ \ \ \ \text{ for}\ R < R_{0} \\
		1 \ \ \ \ \ \ \ \ \ \ \ \ \text{ for}\ R > R_{1} \\
		f(x) \ \ \ \ \ \ \ \text{ for}\ R_{0} \leqslant R \leqslant R_{1}.
	\end{cases}
	\label{eq:pot_trans_coeffs}
\end{split}
\end{align}

In the second case, where the single-star potential dominates in the inner disc, $W_\mathrm{binary}(R)$ takes the values of $W_\mathrm{single}(R)$ from the first case and vice-versa. Using the above prescription the first term in Eq. \ref{eq:pot} is replaced with the following transitioning potential:
\begin{equation}
\Phi_\mathrm{trans} = W_\mathrm{binary} \sum_{k=1}^{2}\Phi_{\mathrm{s},k} + W_\mathrm{single} \Phi_\mathrm{single},
\label{eq:pot_trans}
\end{equation}
where $\Phi_{\mathrm{s},k}$ is the same as that shown previously in Eq. \ref{eq:pot_star} and $\Phi_\mathrm{single} = -\frac{G\left(M_\mathrm{A} + M_\mathrm{B} \right)}{R}$ is the potential created by a single central star of mass $M = M_\mathrm{A} + M_\mathrm{B}$. As can be seen from Eq. \ref{eq:pot_trans_coeffs} $W_\mathrm{b}$ and $W_\mathrm{sing}$ are anti-symmetric so that at 1.5 au, halfway through the transition region, the potential is dominated by neither the binary or single-star potentials.

The results from these two simulations are presented as snapshots of the evolution of the disc surface density profile, in Figs. \ref{fig:sg_pt_grid} and \ref{fig:sg_rpt_grid}. Included are azimuthally-averaged surface density profiles of each snapshot, where perturbations from the background profile can sometimes be more easily seen. The limits of the transition region (1 to 2 au) are plotted in both plot-types. Times for the snapshots have been chosen so that -- where applicable -- we see the birth of any new asymmetric features in the disc and the evolution of the discs to their pseudo-steady-state.

Examining the results from Fig. \ref{fig:sg_pt_grid}, where the binary dominates the potential in the inner disc, we see that the characteristic eccentric central cavity forms as expected. This disc looks almost identical to the Kepler-16 low-mass discs throughout its whole lifetime. The disc reaches pseudo-steady-state quickly, with no additional eccentric features forming in the exterior disc. In contrast, the results from case 2 in Fig. \ref{fig:sg_rpt_grid}, show a range of behaviour. For close to \Pb{2000} the inner disc is axisymmetric, with the gas following circular orbits. Little perturbation by the binary on the outer disc can be seen at this time. In the next snapshot at $t = $\Pb{3000} a large-scale trailing $m=1$ spiral feature can be seen reaching from out beyond the transition region into the inner region. As this feature grows throughout the next few snapshots it perturbs the material in the inner disc. In the transition region, a surface density maximum develops associated with an eccentric ring that forms. The eccentric ring is very similar in structure and evolution to those seen in the high-mass Kepler-16 discs. By combining the final snapshots of the inner region of case 1 with the outer region of case 2, we are essentially able to recreate the results seen for the \mmsn{10} Kepler-16 disc, where the disc sees a potential created by a binary at all positions.

The lack of any additional eccentric rings or spirals in case 1 demonstrates that it is the effect of the binary potential on the self-gravitating disc which excites these features. As the central cavity has formed, along with the synonymous strong surface density maximum, if their creation was due to secular perturbations permitted by self-gravity then we would expect to see them in this case. The lack of the influence of the binary potential in the outer disc is the only ingredient missing from this simulation. The reverse of this can be said about case 2. Here we see an $m=1$ spiral forming at early times in the outer disc, with no prior surface density maximum or asymmetry in the inner disc. This spiral arm always has a pattern speed that is very similar to the precession frequencies of the additional eccentric features that eventually form. This spiral feature only begins to appear in the disc where the binary potential begins to dominate. With no eccentric ring present in the inner disc, there can be no strong secular perturbations on the outer disc exciting these eccentric features. Although this analysis disproves our previous suggestion that secular-like interactions with the inner eccentric disc could lead to the creation of the eccentric features seen in these discs, secular interactions still influence their interaction. In particular, secular forcing could still explain the close-to-commensurate precession frequency ratios seen for the two additional eccentric features seen in the Kepler-16 \mmsn{10}, \mmsn{20} and Kepler-35 \mmsn{20} discs calculated earlier in this analysis.

\begin{figure}
\includegraphics[width = 0.44\textwidth]{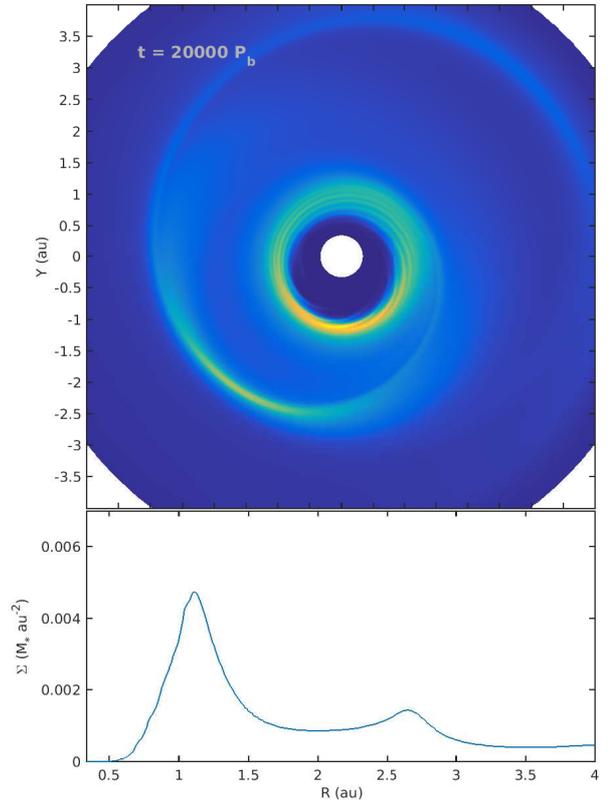}
	\caption{Snapshot of the surface density, and azimuthally averaged profile once steady-state has been reached. The indirect term has been included for the disc evolution. The appearance of the additional eccentric features, and the similarity to our stock self-gravitating disc models, reassures us that they are robust physical feature caused by self-gravity in massive circumbinary discs.}
	\label{fig:sg_idt_grid}
\end{figure}

Fixing the binary orbit as a way of minimising the parameter space, and because of our intent to reproduce the specific Kepler systems simulated here, could produce numerical artefacts in the disc. Whilst simulating a fully "live" binary in these discs is infeasible, because the resulting disc forces will cause the binary orbital elements to diverge from those of the Kepler systems, we set up a test to ensure that the features we see are true physical phenomenon. In our self-gravitating runs above, the effect of the disc on the evolution of the binary was ignored. If this acceleration is included, the centre of mass of the binary is also accelerated, which also acts on the disc as an indirect term when working in a frame based on the binary centre of mass. To simulate a "pseudo-live" binary we calculated the acceleration on each binary member from the disc-binary interaction: $\ddot{\mathbf{r}}_\mathrm{A},\,\ddot{\mathbf{r}}_\mathrm{B}$, and the acceleration on the binary centre-of-mass:
\begin{equation}
	\ddot{\mathbf{r}}_\mathrm{cmbin} = \frac{M_\mathrm{A}\ddot{\mathbf{r}}_\mathrm{A} + M_\mathrm{B}\ddot{\mathbf{r}}_\mathrm{B}}{M_\mathrm{A} + M_\mathrm{B}},
	\label{eq:cmbin}
\end{equation}
which can then be used to obtain an indirect potential throughout the disc, $\Phi_\mathrm{ind} = \ddot{\mathbf{r}}_\mathrm{cmbin}\cdot\mathbf{R}$. Whilst the disc-binary accelerations are not used to evolve the binary orbital elements, an indirect acceleration is included when evolving the disc:
\begin{equation}
	\ddot{\mathbf{r}}_\mathrm{ind} = -\ddfrac{\Phi_\mathrm{ind}}{R} = -\ddot{\mathbf{r}}_\mathrm{cmbin}
\end{equation}
As outlined earlier in the paper, this approach allows the orbital elements of the binary system to remain fixed in time, while allowing the centre of mass of the binary system to move under the gravitational acceleration of the disc. It is equivalent to conserving the centre of mass of the combined disc-binary system while working in a frame based on the binary centre of mass, without allowing the binary orbital elements to change with time. The indirect force creates a uniform acceleration across the whole disc at each time step that involves the disc being accelerated in a direction that is opposite to that experienced by the binary centre of mass. Using this set-up we undertook a test of the Kepler-16 \mmsn{10} model, as above, but with the full binary potential throughout the disc. As expected, we find that including this term has very little influence on the simulation results, as shown in Fig. \ref{fig:sg_idt_grid}. The reason why allowing the binary centre of mass to become live has little influence on the results is because the indirect term acts to accelerate the disc uniformly rather than differentially, such that any features that develop in the disc structure are unlikely to be influenced by this term.

Although we have shown that the additional eccentric features are due to the binary potential acting on the disc at large radius, we have not yet managed to identify the mechanism responsible for creating these. As part of further investigation, we also ran simulations with transitions in the binary potential for the case of circular binaries, and in one case with the binary mass ratio being unity. We also find the growth of additional eccentric features in these simulations, indicating that the central binary does not need to be eccentric for these features to develop. Given that the epicyclic frequency differs in a self-gravitating disc compared to a Keplerian one, we examined whether or not the 3:1 Lindblad resonance could shift outwards in the more massive discs to explain the origin of the first additional eccentric feature, but we find that the change in epicyclic frequency is small and the Lindblad resonance positions only shift by a few percent. The observation of a large scale $m=1$ spiral feature also suggests that coupling at a shifted 3:1 resonance is not responsible for the growth of the eccentric features (coupling at the 3:1 resonance produces a $m=2$ in the original analysis). At the present time the origin of these features remains unexplained, and will be the subject of a future investigation.

\section{Summary and Discussion}\label{sec:sumdis}
We have presented the results of hydrodynamic simulations from several related investigations into disc evolution applied to the Kepler-16, -34 and -35 circumbinary systems. The first of these was a revisiting and comparison of boundary condition choice at the inner disc boundary. Several authors have examined this issue, with no real consensus between outcomes. We have demonstrated, with a side-by-side comparison of the Closed, Open and Viscous outflow conditions in each system, that a wide range of behaviours can be seen. The lack of agreement for each boundary condition between systems, motivated development of a more realistic treatment of the inner disc. Surprisingly, the limiting Viscous condition, which lies between the Open and Closed behaviour, showed the most erratic results, with no clear relation to the Open or Closed model in any of the systems.

By systematically shrinking the inner boundary from the canonical value of $R_\mathrm{in} = 1.5a_\mathrm{b}$ that has been used in previous studies, until the central binary in the Kepler-16 system is completely embedded in the computational domain and the circumbinary disc, we hoped to find a compromise between increased run-time due to a smaller time-step and accurate modelling of mass-flow and angular momentum transfer around and onto the binary. The results from this investigation showed that models converged on the fully-embedded case as the inner boundary shrank, with $R_\mathrm{max}$, the location of the surface density maximum at the outer edge of the tidally cleared cavity, decreasing towards $1.2$ au. The explanation as to why an embedded binary should give rise to a different disc structure than a binary--disc system with a larger inner radius is unclear. In PN13 the role of non-linear mode coupling and the 3:1 outer Lindblad resonance was discussed with reference to the growth of disc eccentricity around circular and eccentric binaries. In the circular case, when the 3:1 outer Lindblad resonance was not contained in the computational domain, disc eccentricity growth was limited. Decreasing the inner disc radius means that a variety of additional resonances are present in the disc. Including these interior resonances may more realistically capture the exchange of angular momentum between the binary and disc, leading to convergence in the size and structure of the tidally truncated cavity which arises as a balance between outward angular flux from the spiral waves excited by the binary and viscous evolution. The smaller cavity size produced by these embedded binary--disc systems is promising for the next stage of our investigation in which we will introduce migrating protoplanets into the evolved self-gravitating discs. As has been previously shown, \citep{Pierens2008, Pierens2008a, Pierens2013, Kley2014, Kley2015}, the inwards migration of these cores is halted near the cavity edge. The larger cavities produced in these past works (using either a large inner Closed, Open or Viscous boundary condition) meant that the final semi-major axes of the planets were larger than their observed Kepler counterparts -- we hope to see the embedded cores migrate closer to their observed positions in our family of disc models, and the results of simulations that examine this will be presented in paper II.

Our updated boundary condition was used in our investigation into the impact of disc self-gravity on the evolution and final structure of circumbinary discs in these systems, where we focus on disc masses that are small enough that the disc is laminar and not in a gravito-turbulent state. We found that in those models with low binary eccentricity (namely the Kepler-16 and -35 systems), the inclusion of self-gravity in low-mass discs had little impact. The clearest difference between disc mass models in each system is the decreased period of disc precession. Examination of Kepler-34, which has a high binary eccentricity, showed that whilst the \mmsn{1} and \mmsn{2} models showed very similar results, at intermediate mass, the inclusion of self-gravity made a significant difference to the final disc structure. At quasi-steady state the \mmsn{5} case had a location for the peak surface density position at the edge of the cavity of $1.9$ au, nearly 0.5 au smaller than the lighter discs. These results highlight again that it is binary eccentricity and not mass-ratio, which has the greatest impact on disc structure. For planets migrating in low-mass discs around binaries with small eccentricity, we would expect them to halt migration at very similar locations given the similarity in disc structure. For more eccentric systems, such as Kepler-34, migration in a \mmsn{5} disc should lead to quite a different outcome compared to lower mass discs. The general effect of disc self-gravity in these systems is to increase the effective mass of the central binary such that radial length scales associated with disc features such as the central cavity become more compact, and high density features such as the surface density maximum associated with the edge of the cavity are more pronounced.

In the high-mass \mmsn{10} and \mmsn{20} models, the impact of self-gravity became very clear. Increasing the disc mass resulted in two very noteworthy changes: (i) significant shrinking of the size of the central eccentric cavity; (ii) formation of additional eccentric ring features in the disc lying outside of the central eccentric cavity. The shrinking cavity size arises because the self-gravity enhances the inward gravitational force that the disc experiences. The additional eccentric features generate pronounced non-axisymmetric features in the disc surface density distributions consisting of crescent-like features (material at apocentre moves more slowly and therefore the surface density is higher there) and a pronounced $m=1$ global spiral wave. The appearance of these additional eccentric rings is most prominent in the high-mass discs around low-eccentricity binaries.

We examined two possible causes for these eccentric features that arise in high-mass discs. The first was resonant secular forcing by the high density eccentric ring that forms at the outer edge of the tidally truncated cavity. This idea, however, can be discounted for two reasons. The first is that material at the outer edge of the inner eccentric cavity precesses at a significantly faster rate than the additional eccentric ring features, and therefore the additional features are not in secular resonance with this material. The second is that customised simulations, where the gravitational potential due to the binary transitioned from that of the binary system to that of a single central star at large disc radii, did not result in additional eccentric features being excited, even though the inner eccentric cavity still formed. The second possible cause for the formation of the additional features was direct coupling to the binary potential at large disc radii. This was tested by using customised simulations where the potential of the central stars transitioned from a single star in the inner disc to a binary system in the outer disc. For a broad range of central binary parameters (mass ratios and eccentricities) we found that the additional eccentric features arose in these simulations, demonstrating that the central binary plays a key role in their excitation. The precise mechanism leading to the growth of these additional eccentric features has not yet been identified, and will be the subject of a future study.

We note that our adoption of a 2-D set-up prevents us from examining potentially important 3-D effects, although these will be examined in a set of forthcoming studies. 
These include the effects of disc warping when the disc and orbit plane of the central binary are misaligned \citep{LarwoodPapaloizou1997}, which can in turn lead to the development of a parametric instability in the disc that may be a source of hydrodynamic turbulence \citep{OgilvieLatter2013}. 
Similarly, the development of eccentric modes in the disc can also lead to parametric instability and hydrodynamic turbulence \citep{Papaloizou2005, BarkerOgilvie2014}. 
A recent study has also shown that a Spiral Wave Instability can also exist in a disc that is tidally forced by a binary system, also leading to a parametric instability and hydrodynamic turbulence \citep{BaeNelsonHartmann2016}. 
Clearly moving to 3-D will allow us to examine these and other effects, increasing the richness of the physical phenomena that can occur in planet forming circumbinary discs. In addition, it will be necessary to examine how changes in physical parameters such as the disc thickness, viscosity and surface density profiles influence the results, not to mention the influence of as-yet neglected physics such as magnetic fields, before we can can make robust claims about the evolution of circumbinary discs.

The significantly different evolution observed for the massive circumbinary discs examined here raise a number of interesting issues concerning the formation and evolution of planets in these environments. In particular, it is clear that the different disc structure would be very important for early forming planets that arise when the disc is massive and self-gravity is important. The surface density maxima produced by the additional excited eccentric rings could act as planet-traps \citep{Masset2006a}, halting planet migration before they reach the edge of the inner cavity. The migrating cores could increase rapidly in mass at these locations due to an accumulation of material \citep{Morbidelli2008}. Although the eccentric features are not vortices (even if they have a similar appearance), they can nonetheless act as pressure bumps and trap small dust grains and pebbles that may grow and migrate through the disc via aerodynamic drag. Their non-axisymmetric structure will place limits on the particle sizes that can be locally trapped (radial migration must be slow enough that particles do not escape by crossing the radial width of these features in one synodic period), but one could certainly expect them to enhance the concentration of solid material locally and provide sites for efficient planet growth. Of course, these high-mass conditions will only occur at very early times in the most massive discs. We therefore need to examine the transition from these high-mass states to more common low-mass ones, and the impact of this transition on the planets which may have formed and migrated at early times. A high-mass past, where the disc is shaped by self-gravity, may leave a lasting imprint on the planets it produces. We will address the question of planetary evolution by investigating migration and accretion scenarios of embedded protoplanets, as well as disc mass transitions, in the forthcoming second paper of this series.

\section*{Acknowledgements}

We acknowledge very useful conversations with John Papaloizou concerning the origin of eccentric ring features in the circumbinary discs, as well as an insightful and constructive report from the referee. Computer time for the simulations performed with {\small GENESIS} was provided by HPC resources of Cines under the allocation c2016046957 made by GENCI (Grand Equipement National de Calcul Intensif). Those performed with {\small FARGO} utilised: Queen Mary's MidPlus computational facilities, supported by QMUL Research-IT and funded by EPSRC grant EP/K000128/1; and the DiRAC Complexity system, operated by the University of Leicester IT Services, which forms part of the STFC DiRAC HPC Facility (\url{www.dirac.ac.uk}). This equipment is funded by BIS National E-Infrastructure capital grant ST/K000373/1 and STFC DiRAC Operations grant ST/K0003259/1. DiRAC is part of the National E-Infrastructure.



\begingroup
\raggedright
\bibliographystyle{mnras}
\bibliography{./sgcbImanuscriptfinal.bib}
\endgroup


\bsp	
\label{lastpage}
\end{document}